# ON THE STATISTICAL NATURE AND DYNAMICS OF SEISMOGENESIS IN THE NW CIRCUM-PACIFIC BELT: A STUDY BASED ON NON-EXTENSIVE STATISTICAL PHYSICS

Andreas Tzanis and Evangeline Tripoliti

*Section of Geophysics,*
*Department of Geology and the Geoenvironment*
*National and Kapodistrian University of Athens*
*Panepistimiopoli, Zografou 15784, Greece*
*E-mail: atzanis@geol.uoa.gr.*

# Abstract

We examine the statistical nature and dynamic expression of shallow (crustal) and deep (sub-crustal) seismogenetic systems along major convergent plate margins of the north-western Circum-Pacific Belt (Ryukyu Izu-Bonin and Honshu Arcs). Specifically, we investigate whether earthquakes therein are generated by self-excited Poisson processes and comprise independent events, or by Complex/Critical processes in which long-range fault-fault interactions render them dependent (*correlated*). The analysis is conducted on the basis of Non Extensive Statistical Physics and searches for evidence of correlation in the earthquake record using the complete and homogeneous earthquake catalogue of the Japan Meteorological Agency for the period 2002-2016.5. Emphasis is given to the analysis of background seismicity, (the rudimentary expression of seismogenetic systems), which is recovered by removing aftershocks (foreground seismicity) with a stochastic declustering procedure.

As a general rule, long-term correlation is moderate in crustal *full* (foreground/background) seismicity and drops to weak in crustal background seismicity; weak to moderate long-range correlation persists in all cases. Conversely, full and background sub-crustal seismicity is long-term uncorrelated (quasi-Poissonian). The time-dependence of correlation is dynamic. Crustal full seismicity reverts between states of very strong (time-local) correlation associated with aftershock sequences and –relaxed– states of moderate correlation in-between. Crustal background correlation is generally weak-moderate but persistent. Sub-crustal full seismicity behaves similarly, but the relaxed-state correlation is generally weak to nihil. Sub-crustal background correlation is generally weak to insignificant but may *dynamically* transition to moderate in association with the occurrence of major earthquakes, for reasons as yet unspecified but presumably related to the geo-dynamic forcing. Crustal and sub-crustal background seismicity evolved from uncorrelated (quasi-equilibrating) to moderately correlated (non-equilibrating) states ahead of the 2011.19 Tōhoku mega-earthquake.

The present results contrast those obtained for the Pacific-North American transformational plate boundaries of California and Alaska, which are strongly correlated; this suggests that the regional geodynamic setting is central to the development of self-organization and complexity. Finally, our results appear consistent with simulations of small-world fault networks in which free boundary conditions at the surface allow for self-organization and possibly criticality to develop, while fixed boundary conditions at depth do not.

## 1. INTRODUCTION

Observed seismicity generally comprises an admixture of two earthquake populations: a *background* that expresses the continuum of regional tectonic deformation and a *foreground* that expresses short-term/high-rate local activity associated with aftershock sequences and earthquake swarms. The collective properties of this *full* (or *whole*) seismogenetic process have conventionally been described with the statistics-based models of statistical seismology. However, over the past two decades a physics-based approach has begun to attract attention. It uses statistical physics in order to bring together physical models of earthquakes genesis and statistical models of earthquake populations, endeavouring to generate statistical models of seismicity from first principles, that respect the laws of thermodynamics and take into account the physical laws of friction, rupture etc. In other words, it uses physics to support stochastic models, a feature often missing from traditional statistical seismology. The discourse between the statistics-based and physics-based description of seismicity has produced two general schools of thought.

The first school postulates that background seismogenesis is Poissonian in spacetime and obeys Boltzmann-Gibbs thermodynamics. This approach is expressed by a series of well-known models that consider background earthquakes to be statistically independent and although it is admissible for one to trigger another, this is assumed to occur randomly in a way that does not influence the independence of successive events and the long-term evolution of the seismogenetic system (fault network). The most influential of these models, is ETAS (Episodic Type Aftershock Sequence, e.g. Ogata, 1998, 1988; Zhuang et al, 2002; Helmstetter and Sornette, 2003; Touati et al, 2009; Segou et al, 2013; many others). Based on the self-excited conditional Poisson process, ETAS posits that randomly occurring earthquakes trigger aftershocks which trigger their own aftershocks, thus producing a series of clustered parental/filial events (aftershock sequences) whose number decays according to the Omori-Utsu power-law (e.g. Utsu et al., 1995). Proxy-ETAS models (Console and Murru, 2001), as well as extended point-process models have been developed in order to address the problem of intermediate to long-earthquake term clustering, such as the PPE (Proximity to Past Earthquakes, e.g. Marzocchi and Lombardi, 2008), the EEPAS (Each Earthquake is a Precursor According to Scale, e.g. Rhoades, 2007).

Poissonian models are mainly concerned with the statistics of time and distance between events. The size (magnitude) distribution of earthquakes is still taken to obey the Gutenberg-Richter frequency–magnitude distribution. This has a significant downside: the Gutenberg-Richter distribution is a power-law that *cannot* be derived from the Boltzmann-Gibbs formalism. Likewise, the Omori-Utsu formula is a Zipf-Mandelbrot power-law and is inconsistent with the Boltzmann-Gibbs formalism. The reliance of Poissonian seismicity models on irrefutable but evidently non-Poissonian empirical laws is a contradiction with no theoretical resolution and demonstrates that they effectively are *ad hoc* conceptual constructs attempting to reconcile the Poissonian worldview of statistical seismology with the apparently non-Poissonian dynamics of faulting and aftershock clustering.

The second school comprises a set of models proposing that earthquakes are generated by complex, non-equilibrating fractal fault networks in which successive earthquakes are *dependent*. The well-known concept of Self-Organized Criticality (SOC) comprises a class of models proposing that the network continuously evolves toward a stationary critical state with no characteristic spatiotemporal scale, in which earthquakes (phase transitions) develop spontaneously and any small event has a chance of escalating to global scale event (e.g. Bak and Tang, 1989; Sornette and Sornette, 1989; Olami et al., 1992; Sornette and Sammis, 1995; Rundle et al., 2000; Bak et al, 2002; Bakar and Tirnakli, 2009; etc.). SOC is self-consistent and also predicts several observed properties of earthquake occurrence: the Gutenberg-Richter law and the Omori-Utsu law emerge *naturally* during the evolution of simulated fault networks. Self-*Organizing* Criticality comprises a class of evolutionary models according to which the network develops a Critical Point at the end of an earthquake cycle (e.g. Sornette and Sammis, 1995; Rundle et al., 2000; Sammis and Sornette, 2002; many others); it was extensively investigated in the late 1990's and early 2000's but is no longer pursued as its basic prediction (acceleration of seismic release rates) could not be statistically verified. In the context of Criticality, the dependence between successive earthquakes (faults) is known as *correlation*, develops *internally*, involves long-range interaction and confers memory (delayed action) to the seismogenetic system; such properties are manifested with power-law statistical distributions of the temporal and spatial dynamic parameters. Finally, a number of authors investigated models with complexity mechanisms that do not involve criticality, yet maintain the fault network in a state of non-equilibrium; for thorough accounts see Sornette (2004) and Sornette and Werner (2009). Notable among these is the Coherent Noise Model (Newman, 1996), which was shown to generate power-law behaviour in the time dependence of successive events (Celikoglu et al., 2010); a drawback is that it does not include the geometric configuration of the "fault network" and it is not known how this would influence its behaviour.

Both of the Poissonian and Complex/Critical schools agree that foreground seismicity comprises dependent events, although the former assigns only local significance and non-critical nature to the dependence, while the latter considers it to be integral part of a regional-scale process. The fundamental difference is in their understanding of the background process, which the former assumes to generate statistically independent earthquake populations (no correlation and memory, exponential distributions), while the latter assumes statistically dependent populations (long-range interaction, power-law distributions). It therefore stands to reason that if it is possible to identify and eliminate the foreground, it would also be possible to study the nature and dynamics of the background by examining its spatiotemporal characteristics for the existence of correlation. In order to pursue this objective, one requires effective measures of correlation and effective methods to separate the background and foreground processes. In addition, one *must* have a self-consistent theoretical framework on which to base the analysis (and not custom models or *ad hoc* conceptual constructs). As explained in Section 2, nearly satisfactory answers exist for all three requirements.

Previous work by Efstathiou et al., (2015, 2016, 2017, 2018) and Tzanis et al., (2013, 2018) has addressed the question of the statistical nature of seismicity by implementing the analytical tools of Non Extensive Statistical Physics in order to study empirical multivariate distributions (joint probabil-ities) of earthquake frequency vs. magnitude and time/distance between successive events which, as will be seen below, comprise effective measures of correlation. These distributions were constructed on the basis of complete and homogeneous earthquake catalogues in which aftershocks were either present (full catalogues), or had been removed with the stochastic declustering method of Zhuang et al., (2002). The previous work was focused on the transformational plate boundaries of the northeastern circum-Pacific belt, (California and Alaska, USA), with only one convergent boundary taken into consideration (Aleutian Arc in Tzanis et al., 2018); it has come up with compelling evidence of strong correlation in the transformational systems and in shallow (crustal) seismicity, but with quite the opposite in the deep (sub-crustal) seismogenetic system of the Aleutian Wadati-Benioff zone. In the present work, the emphasis is shifted to the convergent plate boundaries of the north-western circum-Pacific belt, which are analysed with the same tools. These are the Ryukyu Arc (Philippine Sea–Okinawa/Yangtze plate convergence), the Izu–Bonin Arc (Pacific–Philippine Sea convergence) and the south segment of the Okhotsk Plate (Honshu Arc comprising the Pacific–Okhotsk and Okhotsk–Amur plate convergences). We also consider the predominantly transformational continental domain of south-western Japan (south-western Honshu on the Amurian Plate).

The study areas were chosen not only for the reliable earthquake monitoring services and catalogues, but also because they allow scrutiny into different seismotectonic contexts: lithospheric seismogenesis along a domain of distributed transform faulting, lithospheric seismogenesis along convergent plate margins and sub-lithospheric seismogenesis in major subducting slabs. A comparative analysis of latter two cases is important in view of the significant differences observed by Tzanis et al. (2018) in the level of correlation between crustal and sub-crustal seismogenetic systems and the ideas proposed as to their origin, which include not only environmental factors (cold brittle schizosphere vs. warm, high-pressure subduction zones), but also the major influence of different boundary conditions (free in the lithosphere-atmosphere interface vs. fixed in subduction zones). In consequence, the analysis of the subject matter will proceed by separating *crustal* and *sub-crustal* earthquakes according to the depth of the Mohorovičić discontinuity.

For the sake of brevity, the reader will be referred to Efstathiou et al., (2017, 2018) and Tzanis et al., (2018) whenever specific details are needed. However, in order to preserve the autonomy, completeness and comprehensibility of the present work, all necessary information will be succinctly provided, including the basic theoretical concepts, the fundamentals of the analytical techniques and brief justification of basic choices and assumptions.

## 2. NON-EXTENSIVE APPROACH TO EARTHQUAKE STATISTICS

### 2.1 Multivariate (joint) probability distributions

A definite indicator of interaction (correlation) between faults is the lapse between consecutive earthquakes above a magnitude threshold: this is variably referred to as *interevent time*, *waiting time, calm time*, *recurrence time* etc. Understanding the statistics of earthquake frequency vs. interevent time (F-T) is fundamental in understanding the dynamics of the fault network. Empirical F-T distributions generally exhibit power-law characteristics with fat tails. In the context of classical statistical seismology, they have been modelled by tailed standard statistical distributions reducible to power laws, as for instance are the gamma and Weibull distributions; examples are given in Bak et al., (2002), Davidsen and Goltz (2004), Corral (2004), Martinez et al. (2005), Talbi and Yamazaki (2010) and others. Other researchers proposed *ad hoc* mechanisms for the generation of power laws by a combination of correlated foreground and uncorrelated background processes: this is a statistical physics approach adopted, for example, by Saichev and Sornette (2013), Touati et al. (2009) and Hainzl et al., (2006). At any rate, Molchan (2005) demonstrated that for a stationary point process, if there is a universal distribution of interevent times, then it *has* to be exponential. A second measure of fault interaction is the *hypocentral distance* between consecutive earthquakes above a magnitude threshold (*interevent distance*). The statistics of the frequency vs. interevent distance (F-D) distributions should be related to the *range of interaction*; it has been studied very few researchers and is rather incompletely understood (e.g. Eneva and Pavlis, 1991; Abe and Suzuki, 2003; Batak and Kantz, 2014; Shoenball et al., 2015). A third seldom acknowledged criterion of correlation is the *b* value of the Gutenberg–Richter law which expresses the scaling of the size-space of active faults (fault hierarchy) and conveys information about their spatial distribution and the homogeneity of the domain they occupy. The F-M distribution is *static* and does not confer information about the dynamic expression of a fault network. Nevertheless, it is an undisputable standard against which to compare physical and statistical descriptions of earthquake size distributions and as such will be used herein.

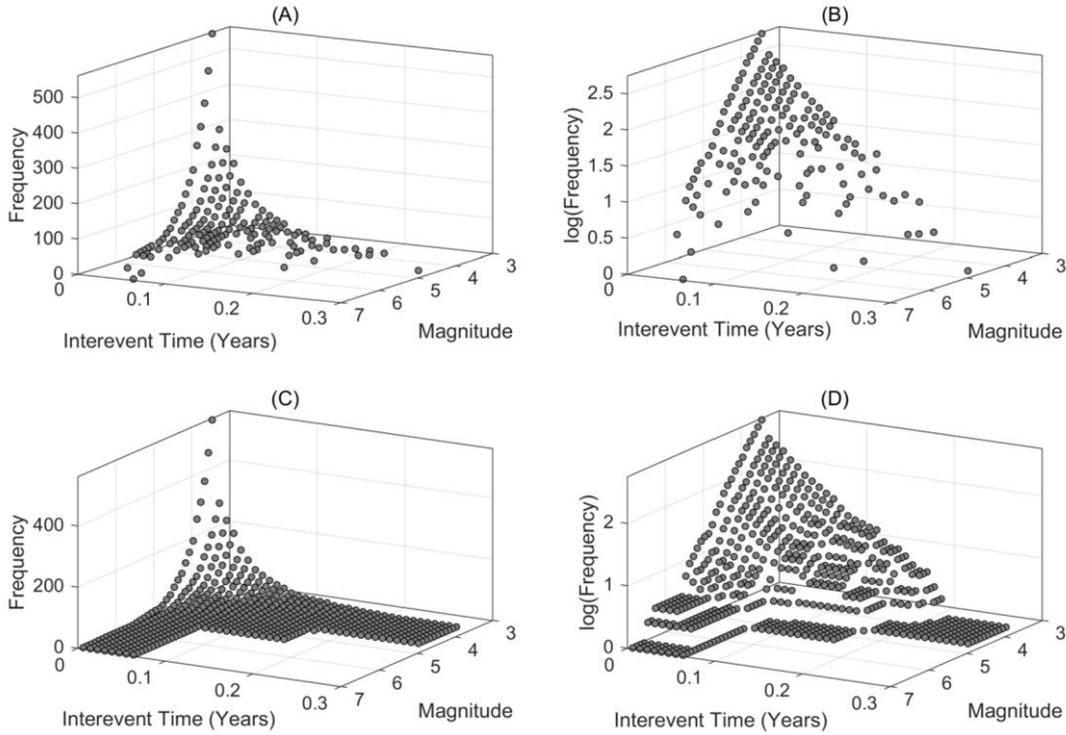

**Figure 1. a)** Bivariate cumulative Frequency–Magnitude–Interevent Time (F-M-T) distribution constructed on the basis of 562 events with $M \geq 3.5$, observed in the continental crust of south-western Japan during 2002-2016.5 (see text for details). **b)** As per (a) but in logarithmic frequency scale. **c)** As per (a) but including un-populated bins in the summation. **d)** As per (c) but in logarithmic frequency scale.

In previous work, (Efstathiou, 2019; Efstathiou et al., 2017, 2018; Tzanis et al., 2018, etc.), the earthquake occurrence model was based on the bivariate joint Frequency–Magnitude–Interevent Time (F-M-T) distribution, which expresses the joint probability of observing an earthquake above of given magnitude and interevent time (within a given range of interevent distances). Adopting the same approach, we construct the F-M-T distribution by setting a threshold magnitude $M_{th}$ and compiling a bivariate frequency table (histogram) representing the *incremental* F-M-T earthquake count (frequency). Letting $H$ denote the empirical incremental distribution, we then compute the empirical *cumulative* earthquake count by backward bivariate summation according to the scheme

$$N_{m\tau} = \sum_{j=D_T}^{\tau} \sum_{i=D_M}^{m} \{H_{ij} \Leftrightarrow H_{ij} \neq 0\},$$

$$\tau = 1, \ldots D_T, \quad m = 1, \ldots D_M$$

where $D_M$ is the dimension of $H$ along the magnitude axis and $D_T$ is the dimension of $H$ along the $\Delta t$ axis. The cumulative frequency (earthquake count) can be written thus: $N(\{M \geq M_{th}, \Delta t : M \geq M_{th}\})$. It follows that, the empirical probability $P(>\{M \geq M_{th}, \Delta t : M \geq M_{th}\})$ is simply

$$\frac{N(>\{M \geq M_{th}, \Delta t : M \geq M_{th}\})}{N(M_{th}, 0)}, \quad (1)$$

$$N(M = M_{th}, 0) = \|N\|_\infty$$

An empirical cumulative F-M-T distribution constructed this way is shown in Fig. 1. It is based on a set of 562 events extracted from the catalogue of the Japan Meteorological agency (JMA) using a threshold magnitude of $M_{th} = 3.5$, which occurred during the period 2002-2016.5 above the Mohoroviciç discontinuity of the continental crust of south-western Japan (see Section 3 for details). The distribution is shown in linear (Fig. 1a) and logarithmic (Fig. 1b) frequency scales and can be seen to comprise a well-defined surface in which the end-member $[M \geq M_{th}, \Delta t = 0]$ is the one-dimensional empirical Gutenberg–Richter law and the end-member [$M=M_{th}$, $\Delta t$] is the one-dimensional Frequency– Interevent Time (F-T) distribution. It is also important to point out that the cumulative distribution is formed by stacking *only* the populated (non-zero) bins of the incremental distribution. If this was not so, the summation would have generated a *stepwise* function in which unpopulated bins lying between populated bins would appear to comprise swathes of equal earthquake frequency (*uniform probability*), as illustrated in Fig. 1c and 1d. In this case, the high probability zones of the distribution would comply with well specified laws, but the lower probability zones would include uniform probability swathes. In one-dimensional distributions, (e.g. empirical G-R distributions), this problem usually has limited effect on parameter estimation. In multivariate distributions however, in addition to the obvious absurdity, it would be numerically detrimental.

Complexity/ Criticality and self-organization are associated with long-range interaction and memory (*correlation*). On the other hand, Poissonian processes are local, independent and memory-less, i.e. uncorrelated. F-M-T distributions are implicitly constructed over the range of interevent distances associated with all pairs of successive events in the earthquake catalogue. Accordingly, in order to study the dependence of correlation on range, it is possible to use the interevent distance as a spatial filter by which to separate and study the temporal correlation of proximal and distal earthquakes; the rationale for doing so is that if distal earthquakes are correlated in time, then they have to be correlated in space by long-distance interaction and vice versa. In consequence, we also analyse F-M-T distributions based on data subsets grouped by interevent distance according to the rule

$C \supset \{C_D: M > M_{th} \wedge \Delta d_L \leq \Delta d \leq \Delta d_U\}$, (2a)

where $C$ is the catalogue, $C_D$ is a subset of the catalogue extracted on the basis of interevent distance, $\Delta d$ is the interevent distance and $\Delta d_L$, $\Delta d_U$ are the upper and lower group limits. This, is equivalent to constructing and modelling the *conditional* bivariate cumulative distribution

$P(>\{M \geq M_{th}, \Delta t : [M \geq M_{th} \wedge \Delta d_L \leq \Delta d \leq \Delta d_U]\})$ (2b)

as a proxy of the *trivariate* F-M-T-D distribution.

## 2.2 NESP formulation of F-M-T joint distributions

In statistical mechanics, an *N*-component dynamic system may have $W=N!/\Pi_i N_i!$ microscopic states, where $i$ ranges over all possible conditions (states).

In classical statistical mechanics, the entropy of that system $S$ is related to the totality of these microscopic states by the Gibbs formula $S=-k\sum_i p_i \ln(p_i)$, where $k$ is the Boltzmann constant and $p_i$ is the probability of each microstate. If the components of the system are all statistically independent (non-interacting), the entropy of the system factorises into the product of $N$ identical terms, one for each component; this is the Boltzmann entropy $S_B=-Nk\sum_i p_i \ln(p_i)$. It is easy to see that one basic property of the Boltzmann-Gibbs entropy is *additivity*: the entropy of the system equals the sum of the entropy of its components. *Additive* systems are also referred to as *extensive* after Tsallis (1988). However, a very broad spectrum of non-equilibrating natural and physical systems includes *statistically dependent* (interacting) components, whose sum of the entropies may *not* be equal to the entropy of the system! Such *non-additive* (*non-extensive*) systems cannot be described with the Boltzmann-Gibbs formalism. They can be described with Non-Extensive Statistical Physics (NESP) which was introduced by Constantino Tsallis as a *direct generalization* of Boltzmann-Gibbs thermodynamics to non-equilibrating systems (Tsallis, 1988, 2001, 2009; Tsallis and Tirnakli, 2010).

In NESP, the non-equilibrium states of some dynamic parameter $x$ can be described by the Tsallis (1988) entropic functional:

$$S_q = k \frac{1}{q-1}\left[1 - \int_W p^q(x)dx\right],  \quad (3)$$

where $p^q(x)dx$ is the probability of finding the value of $x$ in $[x, x+dx]$ so that $\int_W p^q(x)dx = 1$, $k$ is the Boltzmann constant and $q$ is the *entropic index*. For $q=1$, Eq. 3 reduces to the Boltzmann–Gibbs functional $S_{BG} = -k\int_W p(x)\ln(p(x))dx$. The Tsallis entropy is concave and fulfils the *H*-theorem, but is not additive when $q \neq 1$; for example, a mixture of two statistically independent systems $A$ and $B$ satisfies $S_q(A, B) = S_q(A) + S_q(B) + (1-q) S_q(A) S_q(B)$. This property is known as *pseudo-additivity* and is further distinguished into *super-additivity* (*super-extensivity*) if $q<1$, *additivity* when $q=1$ and *sub-additivity* (*sub-extensivity*) if $q>1$. The entropic index is an effective measure of *non-extensivity*.

By maximizing $S_q$, it can be shown that when $q>0$ and $x \in [0, \infty)$, the *cumulative* probability function (CDF) of $x$ is the *q-exponential distribution* (Tsallis, 1988, 2009; Abe and Suzuki, 2005)

$$P(>x) = \exp_q\left(-\frac{x}{x_0}\right) = \left[1-(1-q)\left(\frac{x}{x_0}\right)\right]^{\frac{1}{1-q}} \quad (4)$$

where $x_0$ is a characteristic value (*q-relaxation* value) of $x$ and

$$\exp_q(x) = \begin{cases} (1+(1-q)x)^{\frac{1}{1-q}} & 1+(1-q)x > 0 \\ 0 & 1+(1-q)x \leq 0 \end{cases},$$

is the *q-exponential* function, such that for $q=1$, $\exp_q(x) = e^x$. As can be seen in Eq. 4, for sub-extensive systems with $q>1$, $P(>x)$ is a power-law with tail. For extensive (random) systems with $q=1$, $P(>x)$ is an exponential distribution. Finally, for super-extensive systems with $0<q<1$, $P(>x)$ is a power-law with a cut-off so that $P(>x)=0$ whenever the argument becomes negative; such systems are characterized by bounded correlation radii.

NESP has already been applied to the statistical description of earthquake occurrence with interesting results; for detailed reviews see Efstathiou et al., (2017), Tzanis et al., (2018) and references therein. Because interevent times are real and positive, the CDF expressed by Eq. 4 is the only NESP formulation applicable to the analysis of F-T distributions. The construction of a CDF appropriate for F-M distributions is more complicated in that it has to be derived from Eq. 3 and to relate the energy stored in faults with the measure of the energy released by earthquakes (magnitude) in a physically consistent way. A first principles approach based on "fragment-asperity models" that consider the interaction of fault asperities and fragments filling the space between fault walls, (which is supposed to modulate earthquake triggering), was spearheaded by Sotolongo-Costa and Posadas (2004) and was further developed by Silva et al., (2006) and Telesca (2011, 2012). Such models differ in their assumption of how the energy stored in the asperities and fragments scales with their characteristic linear dimension. We assert that the model proposed by Telesca (2011, 2012), which assumes that the energy scales with the area of the fragments and asperities ($E \propto r^2$) so that $M \propto \frac{2}{3}\log(E)$, is consistent with the empirical laws of energy–moment and moment–magnitude scaling and is also compatible with the well-studied rate-and-state friction laws of rock failure. Accordingly, the F-M distribution used herein is

$$P(>M) = \frac{N(>M)}{N_0} = \left(1 - \frac{1-q_M}{2-q_M} \cdot \frac{10^{2M}}{\alpha^{2/3}}\right)^{\left(\frac{2-q_M}{1-q_M}\right)}, \quad (5)$$

with the constant $\alpha$ expressing the proportionality between the released energy $E$ and the fragment size $r$ and $q_M$ is the *magnitude entropic index*.

The distributions of magnitude $M$ and interevent time $\Delta t$ are generated by independent processes so that the joint probability $P(M \cup \Delta t)$ can factorize into the probabilities of $M$ and $\Delta t$:

$$P(M \cup \Delta t) = P(M) P(\Delta t). \quad (6)$$

On combining equations (1), (4), (5) and (6), removing the normalization and taking the logarithm one obtains

$$\log N(>\{M \geq M, \Delta t : M \geq M\}) = \log(N_{M=0})$$
$$+ \left(\frac{2-q_M}{1-q_M}\right) \cdot \log\left(1 - \frac{1-q_M}{2-q_M} \cdot \frac{10^M}{\alpha^{2/3}}\right) \quad (7)$$
$$+ \frac{1}{1-q_T}\log\left(1 - \Delta t_0^{-1}(1-q_T)\Delta t\right)$$

where $q_M$, $q_T$ are entropic indices for the magnitude and interevent times respectively and $\Delta t_0$, is the *q-relaxation time*, analogous to the relaxation time often encountered in the analysis of physical systems. Eq. 7 is a generalized (bivariate) Gutenberg – Richter law in which

$$b_q = \frac{(2-q_M)}{(q_M-1)} \quad (8)$$

is the NESP *generalization* of the $b$ value.

The parameters of Eq. 7 can be approximated with non-linear least-squares. Because they are all positive (bounded from below), and the entropic indices are also bounded from above, we have chosen to implement the *trust-region reflective* algorithm (e.g. Moré and Sorensen, 1983; Steihaug, 1983), together with *least absolute residual* (LAR) minimization to suppress outliers. Fig. 2a illustrates the model fitted to the data of Fig. 1b. The solution is associated with 557 degrees of freedom and the approximation (continuous surface) is excellent (correlation coefficient $R^2 \approx 0.987$). The magnitude entropic index $q_M = 1.521\pm0.001$ so that $b_q \approx 0.92$, which compares well to $b$ values of 0.88 to 0.96 computed for the same data set with conventional techniques. The temporal entropic index $q_T$ is approximately $1.362\pm0.02$ and indicates moderate sub-extensivity. $(\Delta t_0)^{-1}$ is $51.24\pm1.13$, meaning that the q-relaxation time (characteristic time for the occurrence of events with $M \geq 3.5$) is 0.0195 years or 7.12 days. Finally, the energy scaling constant $\alpha = 0.0646\pm0.0033$.

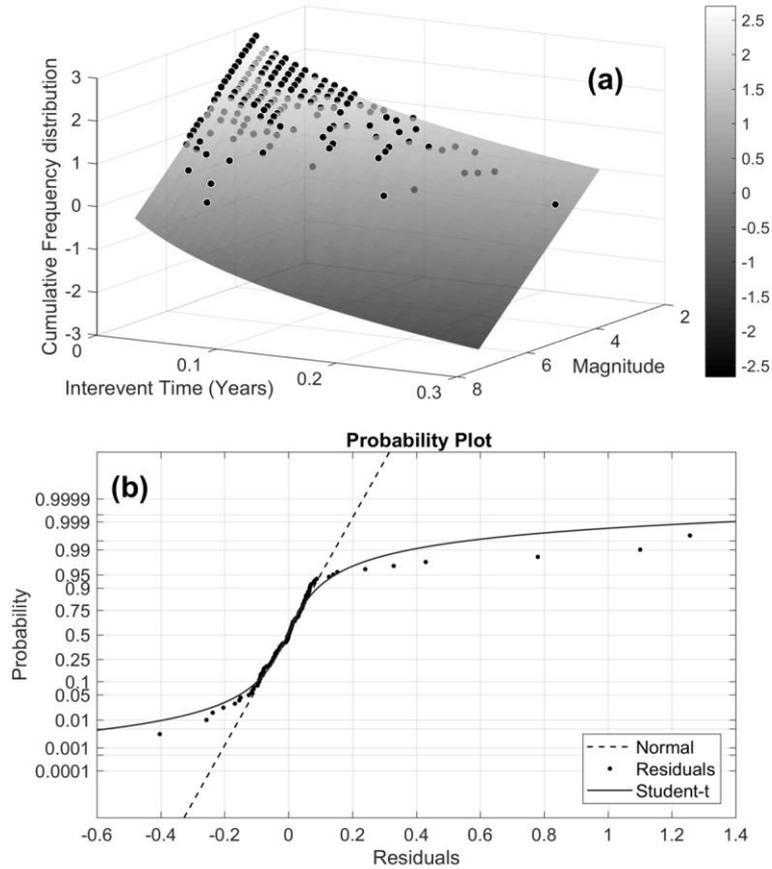

**Figure 2. (a)** The logarithm of an empirical F-M-T distribution (solid circles) and a model fitted using Eq. (7) and the Least Absolute Residual minimisation (transparent surface). **(b)** Probability analysis of the residuals computed by subtracting the model from the empirical F-M-T distribution (see text for details).

Fig. 2b presents a succinct statistical appraisal of the result, performed by fitting a normal location-scale distribution (dashed line) and a Student-t location-scale distribution (solid line) to the cumulative probability of the sorted residuals ($r$). Approximately 97.5% of the residual population, for which $-0.125 \leq r \leq 0.125$, is normally distributed. The tails forming at $r < -0.125$ and $r > 0.125$ are fairly well fitted with the *t*-location-scale distribution and thus represent statistically expected outliers (effectively suppressed by the LAR procedure). It is interesting to note that the properties of the distribution are determined by the populous small-intermediate magnitude scales and interevent times, and that outliers are mainly observed at the intermediate-large magnitude scales and longer interevent times. Outliers frequently arise from minor flaws of the catalogue, (e.g. omitted events, glitches in magnitude reporting etc.), but in some cases could be genuine exceptions to the norm of the regional seismogenetic process: for instance, they may correspond to rare, externally triggered events. We shall not be concerned with such details but it is nonetheless interesting to point them out.

When this type of analysis is carried out for different magnitude thresholds and interevent distance groups, one obtains tables and graphs of the variation of the entropic indices, $q$-relaxation time, energy scaling constant and other parameters, which provide a concise description of the statistical properties of seismicity. Fig. 3 illustrates the analysis of a catalogue of 1763 events with $M \geq 3.0$, which occurred in the continental crust of southwestern Japan during 2002-2016.5 (extracted from the JMA catalogue – see Section 3 for details). The variation of the entropic indices with magnitude is shown in Fig. 3a; they will be thoroughly examined in Section 4.3 and will not be elaborated herein. Fig. 3b illustrates the variation of $1/\Delta t_0$ (inverse of the *q*-relaxation time) and Fig. 3c the variation of the energy scaling constant $\alpha$. It is interesting to observe that $1/\Delta t_0$ decreases exponentially with threshold magnitude, as can be verified by fitting an exponential decay model; this means that $\Delta t_0$, i.e. the characteristic time scale for the occurrence of earthquakes with magnitude $M \geq M_{th}$, increases exponentially with magnitude. Finally, Fig. 3d and 3e respectively illustrate the variation of the goodness of fit ($R^2$) and degrees of freedom associated with the numerical solution of Eq. 7. In what follows, we shall focus our analysis on the magnitude ($q_M$) and temporal ($q_T$) entropic indices, which summarize the (dis)equilibrium states of seismogenetic systems. The significance and possible utility of the other parameters ($\Delta t_0$, $\alpha$) will be examined in future work.

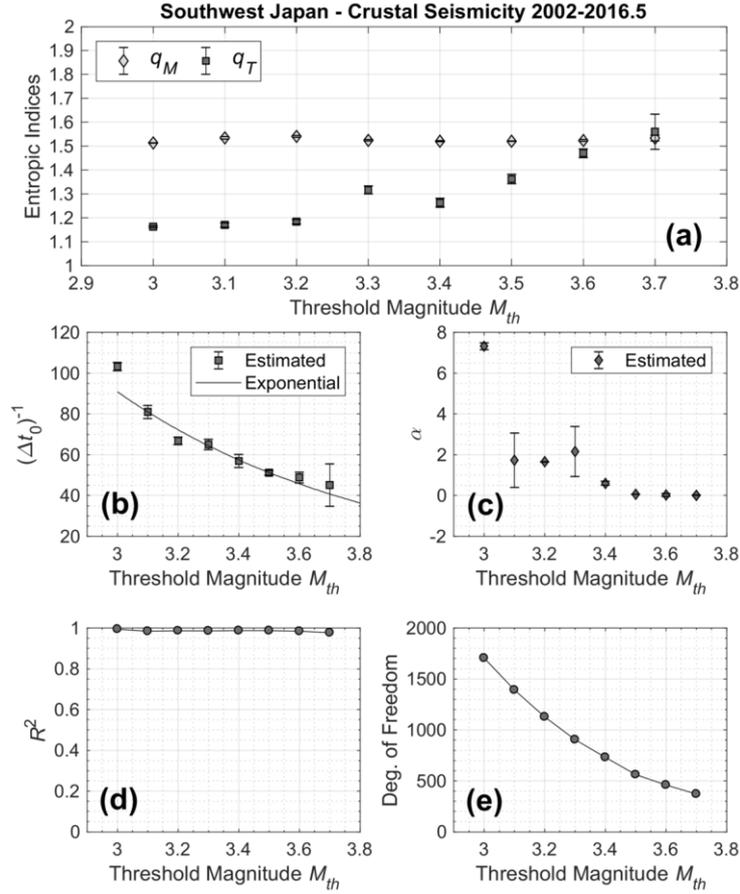

**Figure 3.** Complete analysis of a catalogue of 1763 events with $M \geq 3.0$, observed in the in the continental crust of south-western Japan during 2002-2016.5 (see text for details). All panels illustrate the variation with threshold magnitude of: **a)** the magnitude ($q_M$) and temporal ($q_T$) entropic indices, **b)** the $q$-relaxation time $\Delta t_0$, **c)** the energy scaling constant $\alpha$, **d)** the goodness of fit ($R^2$), and, **e)** the degrees of freedom associated with the numerical solution of Eq. 7.

## 3. EARTHQUAKE SOURCE AREAS

Our study focuses on a broad area of the NW circum-Pacific belt, from approx. 22°N to 46°N (Hokkaido Island, Japan) and from 122°E (east of Taiwan) to 146°E in the Pacific Ocean. As can be seen in Fig. 4, this includes several major convergent and one divergent plate boundaries, transformational plate boundaries and inland seismogenet-ic domains. Of these, the divergent, transformational and inland seismogenetic systems are mainly *crustal*: earthquakes occur mostly in the *schizosphere* (i.e. the rigid, brittle part of the upper lithosphere). On the other hand, the convergent systems are both crustal and *sub-crustal*. As we are interested in testing for differences in crustal and sub-crustal seismicity, we have chosen to examine crustal and sub-crustal earthquake populations independently by separating them according to the local depth of the Mohorovičić discontinuity.

The Mohoroviciç discontinuity and upper mantle structures in and around the Japanese territories has been well investigated with active and passive seismic studies (e.g. Iwasaki et al., 1990; Yoshii, 1994; Iwasaki et al., 2002; Nakamura et al., 2003; Yoshimoto et al., 2004; Hasegawa et al., 2005; Shiomi et al., 2006; Nakamura and Umedu, 2009; Chou et al., 2009 and Uchida et al., 2010). Analogous data for the broader study also exist in the 1°×1° global crustal model of Laske et al., (2013)[1]. Accordingly, we assembled information about the depth to the discontinuity from all sources quoted above and interpolated it onto 0.1°×0.1° model of the Moho surface across the entire study area. This is illustrated in Fig. 5 and is the basis for separating crustal and sub-crustal seismicities.

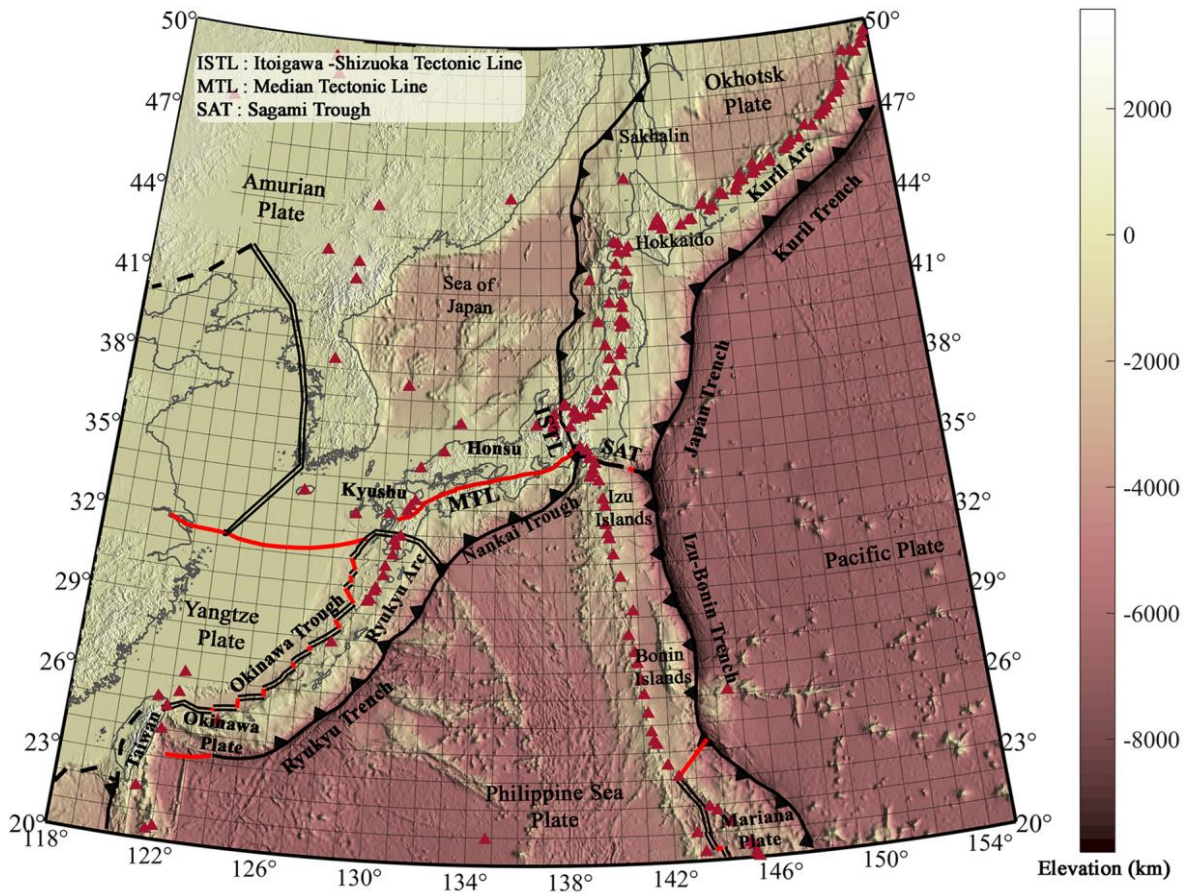

**Figure 4:** The geodynamic setting of the study area. Solid black triangles indicate convergent plate margins (large for rapid, small for slow convergence); double lines indicate divergent plate margins; thick solid lines transformational plate margins and dashed lines "other" margins. Up triangles indicate active volcanoes. The plate boundary data was extracted from Bird (2003) and the elevation data from the ETOPO1 database (Amante and Eakins, 2009).

The earthquake data used in this study span the period 01/01/2002–31/05/2016 and was extracted from the catalogue of the Japan Meteorological agency (JMA), compiled on the basis of information provided by several Japanese research and educational agencies and made available by the National Research Institute for Earth Science and Disaster Resilience (NIED) of Japan[2] (see Acknowledgements for details). The JMA catalogue is homogeneous by construction and complete for $M≥3.0$ (e.g. Fig. 7). The epicentres of shallow/ crustal earthquakes are shown in Fig. 5, on top of the Mohorovicíç discontinuity model used for separating them. The hypocentres of sub-crustal/deep earthquakes are illustrated in Fig. 6. Only earthquakes with magnitudes $M≥3$ are included in the analysis; this is a compromise adopted so as to ensure completeness, sufficiently large earthquake populations to guarantee statistical rigour and exclusion of distant small events that are ostensibly uncorrelated and may randomize the statistics (for more details see Sections 4 and 5).

The statistical nature of background seismogenesis may be examined by analysing reduced versions of earthquake catalogues in which aftershock sequences are eliminated in an optimal as possible way. The process of separating background and foreground events is referred to as *declustering*. We have chosen to implement the declustering method of Zhuang et al. (2002) because it has significant for our objectives advantages. i) It *optimizes* the temporal and spatial window in which to search for aftershocks by fitting an ETAS model to the earthquake data. ii) It assigns each earthquake with a probability that it is aftershock of its predecessor, so that all earthquakes may be possible main shocks to their short-term aftermath. iii) It is a *paradigmatic* realization of the self-excited Poisson process so that if background seismicity obeys Boltzmann-Gibbs statistics, the method should be able to extract a random earthquake ensemble; if not, the argument in favour of non-Poissonian background seismogenesis becomes stronger. The Zhuang algorithm assigns a probability $\phi_j$ for the $j^{th}$ event in the catalogue to belong to the background and thus separates the catalogue into a series of sub-processes whose *initiating events* comprise the background. As a rule of thumb, events with $\phi_j \leq 50\%$ are taken to be foreground while events with $\phi_j > 50\%$ are likely to be background. The results of the declustering exercise are summarized in Table 1 and illustrated in Fig. 8 (crustal seismicity) and Fig. 9 (sub-crustal seismicity), in the form of cumulative earthquakes counts of the *full* (mixed background/foreground) and declustered catalogue realizations. In all cases the *full process* counts are shown as solid lines and the *declustered* at the $\phi_j \geq 70\%$, $\phi_j \geq 80\%$ and $\phi_j \geq 90\%$ probability levels with dashed, dash-dot and dotted lines respectively. Additional information can be found in Efstathiou et al., (2017).

---
[1] http://igppweb.ucsd.edu/~gabi/rem.html
[2] http://www.hinet.bosai.go.jp

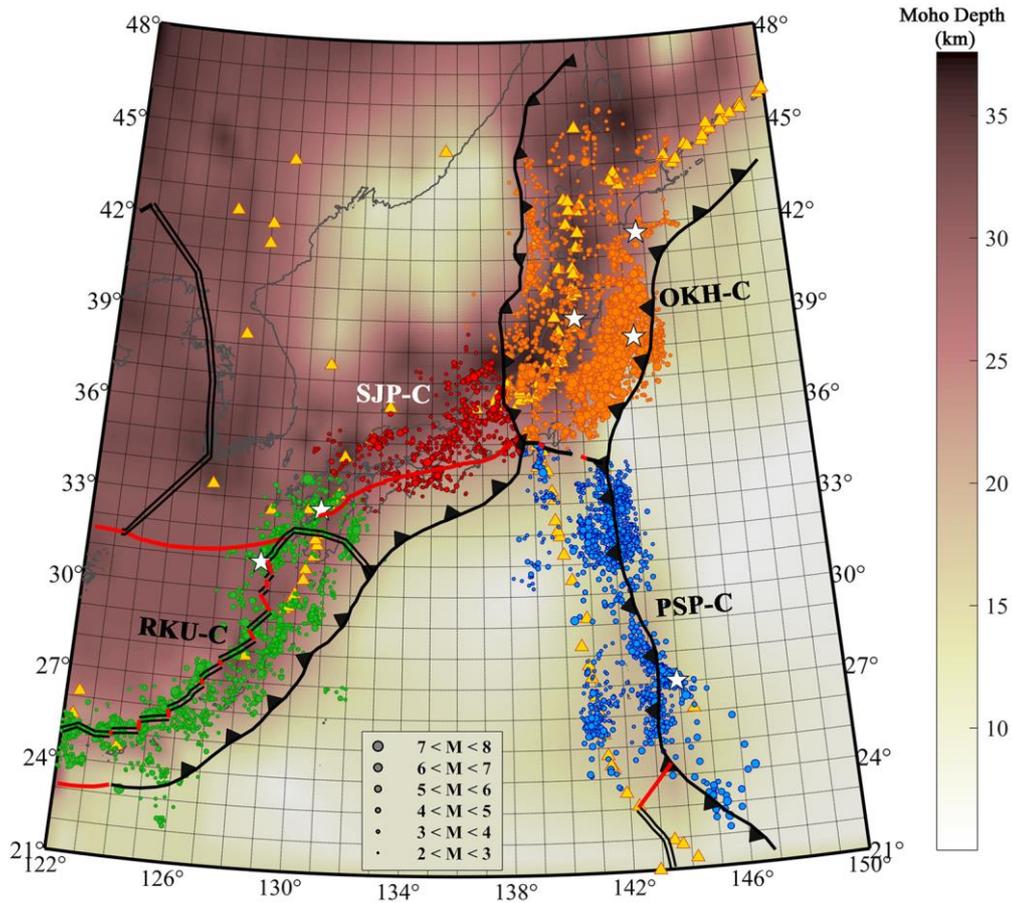

**Figure 5:** The topography of the Mohoroviçiç discontinuity (shaded relief image) and the epicentres of *crustal* (above Moho) earthquakes with magnitudes *M*≥3 observed in the study area during 2002-2016.5. Plate boundaries are drawn as per Fig. 4 and up triangles again indicate active volcanoes.

The seismogenetic systems and fault networks examined herein are as follows:

**a) RKU:** The divergent Yangtze – Okinawa plate margin (Okinawa Trough) and the convergent Okinawa – Philippine Sea plate margin, which forms the Ryukyu Trench and Arc. These run parallel to each other roughly between (123E°, 23°N) and (132°E, 33°N) in Kyushu Island, Japan and form an arcuate system bulging to the southeast. The Ryukyu Trench marks the subduction of the oceanic Philippine Sea Plate beneath the continental Okinawa and Yangtze plates, which occurs at an average rate of 52 mm/yr. The Ryukyu Arc is a ridge comprising two parallel chains of more than 100 islands, with those of the inner arc being Quaternary island arc volcanoes created by the subduction of the Philippine Sea Plate, and those of the outer arc being non-volcanic (Iwasaki et al., 1990). The Okinawa Trough is a rift structure comprising the back-arc basin of the Ryukyu Trench–Arc–Back Arc system (Lee et al., 1980; Kobayashi, 1985; Sibuet et al., 1987).

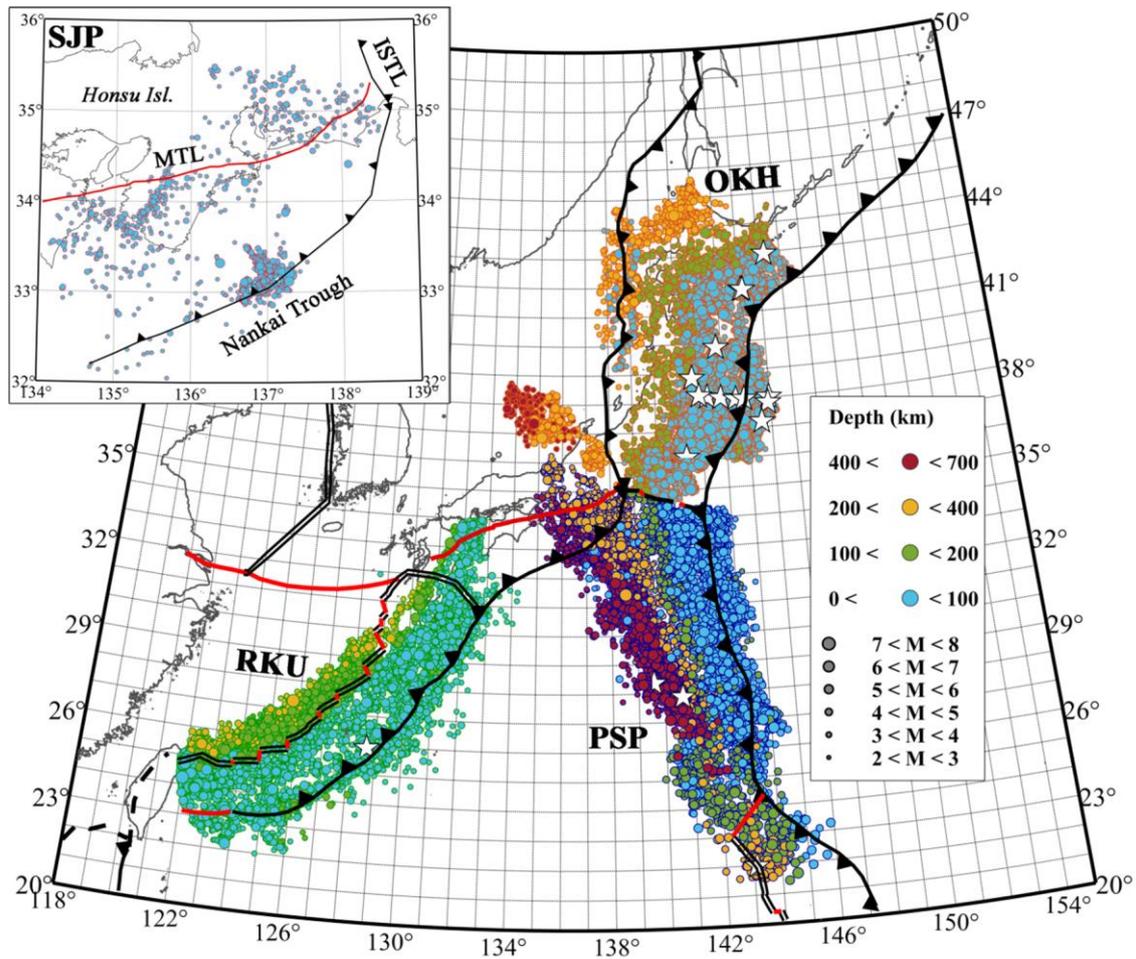

**Figure 6:** Epicentres of *sub-crustal* (below Moho) earthquakes with magnitudes $M \geq 3$ observed in the study area during 2002-2016.5. Plate boundaries are drawn as per Fig. 4.

The (crustal *and* sub-crustal) RKU catalogue is homogeneous by construction and complete for $M \geq 3.0$ (Fig. 7a). Shallow earthquakes are highly clustered and occur mostly in areas with crustal thickness greater than 20km, i.e. continental crust (Fig. 5); they are apparently aligned with the Okinawa Trough where they presumably are tectonic, and with the Ryukyu Island Arc where they presumably are tectonic and volcano-tectonic. During the period 2002-2016.5, crustal seismicity was characterized by a series of large to major earthquakes (2002/3/27 M7; 2005/3/20 M7; 2007/4/20 M6.7; 2015/11/14 M7.1; 2016/4/16 M7.3) which with the exception of the event, they all had *low-intensity*, *short-lived* aftershock sequences (Fig. 8a). Declustering has effectively removed the aftershock sequences: at the $\phi \geq 70\%$ probability level the catalogue is *almost* free of time-local jerks indicative of their presence and any small portion of leftover foreground is progressively eliminated at higher probability levels (Fig. 8a).

As evident in Fig. 6, sub-crustal seismicity is more or less evenly distributed over the subducting slab. Between the Ryukyu Trench and the Okinawa–Yangtze plate boundary, focal depths are mainly concentrated directly below the Trench and are confined to depths shallower than 100km; they plunge abruptly into the mantle just behind the Okinawa–Yangtze boundary and beneath the Okinawa Trough, to depths no greater than 300km. Sub-crustal seismicity is characterized by numerous (15) large to major events, the strongest of which (M7.2) occurred on 2010/2/27; they were all associated with very low-intensity and very short-lived aftershock sequences (Fig. 9a). The quasi-linear nature of the full sub-crustal earthquake count, as well as of its declustered realizations, also show that sub-crustal seismicity along the RKU comprises a larger proportion of background events associated with a proportionally small number of foreground events.

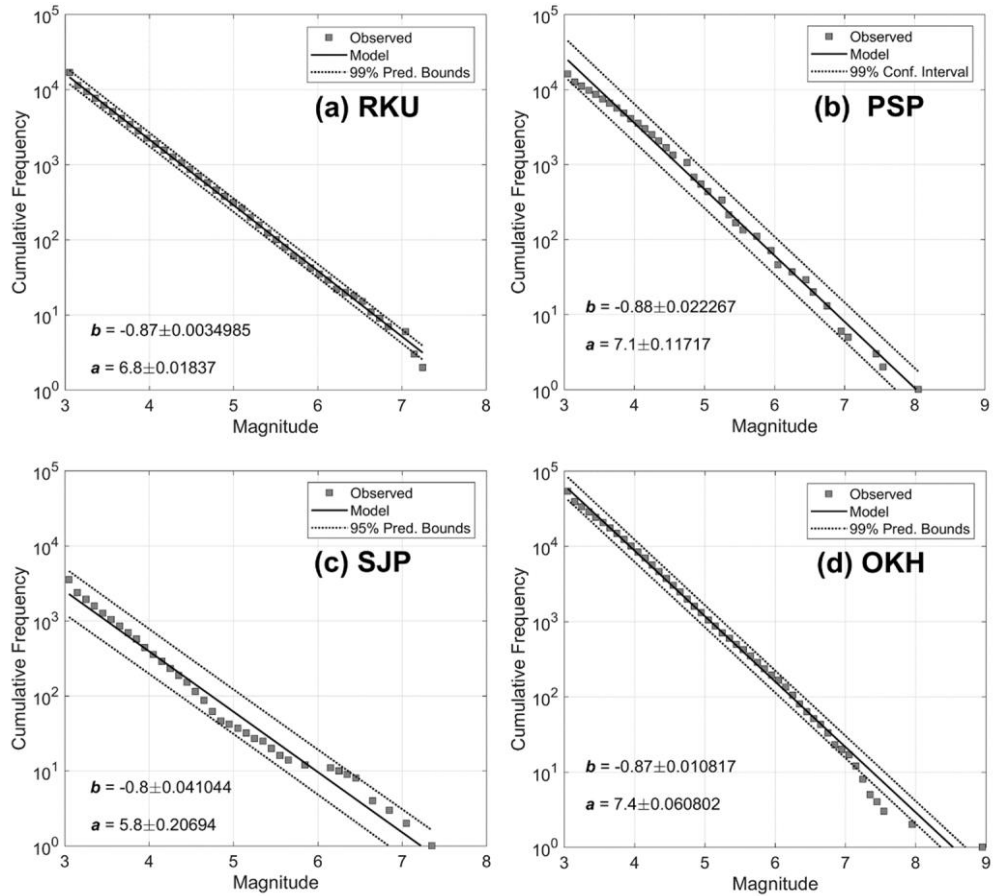

**Figure 7:** The Frequency – Magnitude distribution of (crustal *and* sub-crustal) seismicity in each of the four major earthquake source areas considered in the present work.

**b) PSP:** The Philippine Sea–Pacific intra-oceanic convergent plate margin, which forms the Izu–Bonin–Mariana (or Izu–Ogasawara–Mariana) Arc and Trench. Herein we concentrate on the 1400km long Izu–Bonin segment, northward of latitude 21°N in the northern Mariana Plate and up to the interface of the Philippine Sea, Okhotsk and Pacific plates at the Boso TTT triple junction (roughly 141.9°E, 34.2°N). The Izu segment is punctuated by inter–arc rifts (Klaus et al., 1992; Taylor et al., 1991) and farther south also contains several submarine felsic calderas (Yuasa and Nohara, 1992). The Bonin segment contains mostly submarine and a few island volcanoes. Crustal thickness averages to 20-22 km with a felsic middle crust. Subduction rates vary from 46 mm/yr in the north to ~34mm/yr in the south (Bird, 2003; Stern et al., 2004). The Wadati-Benioff zone varies along strike, from dipping gently and failing to penetrate the 660 km discontinuity in the north, to plunging vertically into the mantle but failing to penetrate the 410km transition in the south (Fig. 6; also see Stern et al., 2004). The north boundary of this system is the 340-kilometre long Sagami Trough (SAT), extending from the Boso triple junction in the east, to Sagami Bay, Japan, in the west. It comprises the surface expression of the convergent boundary, along which the Izu forearc of the Philippine Sea Plate is subducted under the Honshu forearc of the Okhotsk Plate (Nakamura et al., 1984; Ogawa et al., 1989). The SAT is known for major ($M \geq 8$) mega-thrust historical and instrumental era earthquakes, as for instance the M7.9 Great Kanto Earthquake of 1923/9/1 (e.g. Kobayashi and Koketsu, 2005).

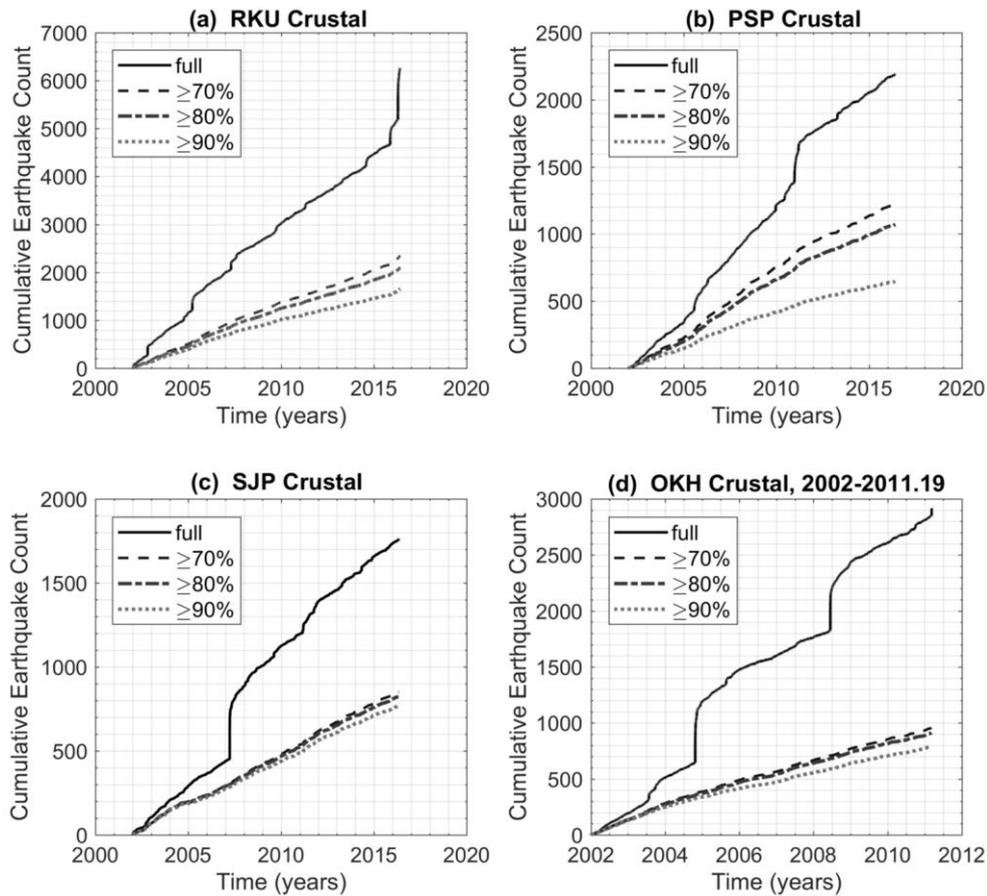

**Figure 8:** The cumulative event counts of full and declustered *crustal* earthquake catalogues in each of the four major earthquake source areas considered in the present work.

The (crustal *and* sub-crustal) PSP catalogue is homogeneous by construction and complete for $M > 3.0$ (Fig. 7b). During the period 2002-2016.5, crustal seismicity has occurred in the Izu-Bonin trench and along the Bonin Volcano Island chain where it is presumably volcano-tectonic (Fig. 5). Only three significant ($M \geq 6.5$) earthquakes have taken place, two of which (2005/2/10, M6.5 and 2006/10/24, M6.8) have low intensity, short-lived aftershock sequences and one (2010/12/22, M7.4) has a prolific sequence (Fig. 8b). Otherwise, earthquake activity appears to be distributed and continuous. It is, thus, hardly surprising that more than half of the crustal events are classified as "background" at the $\phi \geq 70\%$ probability level, and approximately half at the $\phi \geq 80\%$ level (Fig. 8b). Notable also is an apparent decrease in earthquake production after 2011, which we shall not investigate herein.

As evident in Fig. 6, sub-crustal seismicity is rather evenly distributed in the subducting slab. Eleven significant ($6.5 \leq M \leq 7$) and four major ($M > 7$) events have taken place during 2002-2016.5, the most noteworthy of which have occurred on 2007/9/28 (M7.6) and 2015/5/30 (M8.1). They all had very low-intensity and very short-lived aftershock sequences, with earthquake activity appearing to be otherwise continuous (Fig. 9b). Accordingly, almost 68% of the sub-crustal earthquakes have been classified as "background" at the 70% probability level, 59% at the 80% level and 37% at the 90% level (Fig. 9b).

**c) SJP:** The geological domain of southwestern Japan consists of the Shikoku and southern half of Honsu islands and extends between Kyushu and the Itoigawa-Shizuoka Tectonic Line (ISTL). This area is part of the Amurian continental crust. Inland crustal seismicity is dominated by the WSW-ENE right-lateral Median Tectonic Line (e.g. Tsutsumi and Okada, 1996) and the Niigata–Kobe Tectonic Zone (NKTZ) which in SW Honshu comprises a dense network of conjugate NW-SE and NE-SW strike-slip systems; the latter is bounded to the south by the MTL and can be explained by an E-W compressional stress regime (e.g. Taira, 2001; Sagiya et al., 2000). The west-ward extension of the MTL connects with a zone of north-south extension in central Kyushu (e.g. Okamura et al., 1992), which continues to the eastern end of the Okinawa trough. The MTL and NKTZ are part of the broad active boundary between the Philippine Sea and Amurian plates. These two plates converge along the Nankai Trough, off the coast of south-western Japan and generate significant intermediate depth seismicity. Convergence directions and rates are N–NW and 4.5cm/yr respectively (Seno et al., 1993). The tectonic coupling between the overriding and subducted plates has been evaluated to nearly 100% over the landward slope of the entire Nankai Trough (complete fault locking, Mazzotti et al., 2000). Several major earthquakes are known to have occurred along the Nankai mega-thrust, (interface between the two plates), with a recurrence period of one to two hundred years (e.g. Cummins et al., 2001 and references therein). The western boundary of SJP is the Itoigawa-Shizuoka Tectonic Line (ISTL) in central Japan, which is part of the convergent front of the Amurian and Okhotsk plates. The ISTL is about 250 km long, several km wide and exhibits long-term slip rates of around 8-10 mm/yr (e.g. Okumura et al., 1994; 1998) and long earthquake recurrence intervals (e.g., Okumura, 2001). Sagiya et al., (2002) determined E-W horizontal shortening of the order of 0.3μstrain/yr and Ikeda et al., (2004) indicated that ISTL comprises a low-angle, low-strength fault plane. Northward of Hoshu, at the eastern margin of the Sea of Japan, the ISTL transitions into the fold-thrust belt of the Amurian–Okhotsk plate boundary (Yeats, 2012).

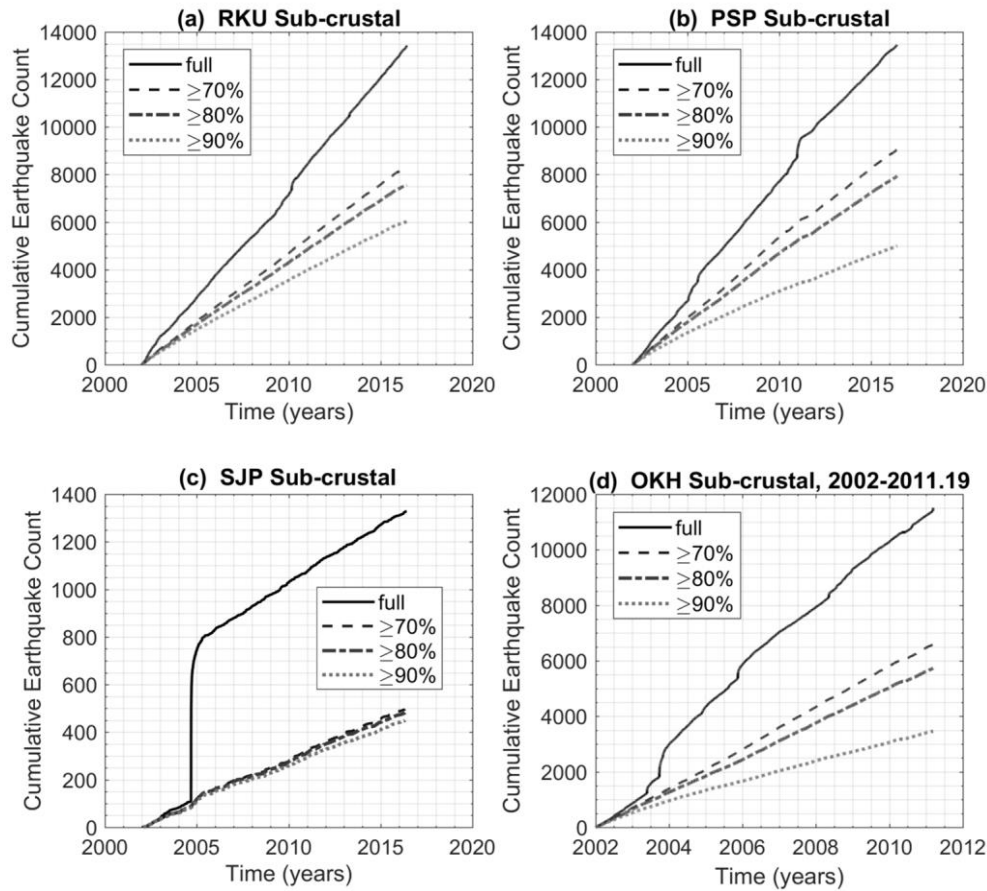

**Figure 9:** The cumulative event counts of full and declustered *sub-crustal* earthquake catalogues in each of the four major earthquake source areas considered in the present work.

The (crustal *and* sub-crustal) SJP catalogue is homogeneous by construction and appears to be complete for $M \geq 3.0$, although the cumulative frequency-magnitude curve exhibits a peculiar oscillation that we shall not attempt explain (Fig. 7c). In the period 2002-2016.5, earthquake activity has been intense but not particularly prolific. Crustal seismicity has mostly been concentrated in the NKTZ and to a considerably lesser degree along the MTL (Fig. 6). Only two large ($M \geq 6.5$) earthquakes have taken place, with the most significant of them (2007/3/25, M6.9) accompanied by an extended aftershock sequence (Fig. 8c); activity has otherwise been limited to a series of distributed intermediate-sized events accompanied by low-intensity aftershock sequences. Foreground events appear to have been successfully removed by the stochastic declustering procedure (Fig. 8c). The sub-crustal activity has generated two major events 37-45 km beneath the Nankai Trough, presumably associated with the Tonankai segment of the Nankai mega-thrust (2004/9/5, M7.1 and M7.4). These were accompanied by a prolific albeit short-lived aftershock sequence that included an additional two M6.5 events (on 7 and 8 of September 2004). Sub-crustal activity has otherwise occurred mostly beneath south-western Honshu and has been continuous, distributed and practically free of foreground events (compare cumulative earthquake counts in Fig. 9c).

### d) OKH:

The Okhotsk Plate and plate boundary, defined to the south by the Sagami Trough, (see above), to the west by the "slowly" (~10 mm/yr) converging fold-thrust front of the Amurian and Okhotsk plates (e.g. Taira, 2001; Yeats, 2012), and to the east by the rapidly converging (~90 mm/yr) boundary of the Okhotsk and Pacific plates. The latter form a subduction zone in which the Pacific Plate is being subducted underneath the Okhotsk Plate; it forms the Japan Trench that extends from the Boso triple junction near (142°E, 34°N) to approximately (145°E, 41°N) and is responsible for creating the broad island arc of North-eastern Honshu. The Japan Trench is succeeded by the Kuril–Kamchatka Arc and Trench system that extends up to the triple junction with the Ulakhan Fault and the terminus of the Aleutian Arc and Trench, near (164°E, 56°N).

The Okhotsk Plate and plate boundary systems have experienced particularly intense activity during 2002-2016.5. Herein we will only consider seismicity data up to the line (146.8°E, 42.5°N) – (140.2°E 46.5°N), north of Hokkaido, Japan. Farther north, a very large part of the JMA catalogue does not contain reliable focal depth information and is neither possible to separate crustal and sub-crustal earthquakes, nor calculate interevent distances. The Amurian–Okhotsk boundary, although responsible for many strong earthquakes in the Sea of Japan and Sakhalin (e.g. Arefiev et al., 2006), has not been particularly energetic. Instead, activity concentrated mainly along the Pacific–Okhotsk subduction, in which many strong mega-thrust earthquakes have taken place such as the 2003 M8.3 Hokkaido event (e.g. Watanabe at al., 2006) and the M9.0 2011 Tōhoku mega-earthquake (e.g.

Ozawa et al., 2011). Significant, albeit not as intense crustal activity has also taken place along the Honshu arc.

As before, the combined crustal *and* sub-crustal OKH catalogue is homogeneous by construction and complete for $M \geq 3.0$ (Fig. 7d). Earthquake productivity has been prolific: a remarkable sixty four $M \geq 6.5$ earthquakes occurred during 2002-2016.5, twenty two of which prior to the 2001.19 Tōhoku mega-event and ten of which were major ($M \geq 7$); the remaining forty two mostly occurred as part of the Tōhoku aftershock sequence, which included a total of nine major events. Crustal seismicity concentrated mainly along the Pacific-Okhotsk forearc and was dominated by the Tōhoku aftershock sequence (Fig. 5), although it has been also been significant along the Honshu Arc and backarc belts. This included twenty four $M \geq 6.5$ earthquakes, seven of which occurred prior to the Tōhoku mega-event and five were major (2003/9/26, M7.1; 2004/10/23, M6.8; 2008/6/14, M7.2; 2011/3/9, M7.3; 2011/3/10, M6.8). The Wadati-Benioff zone dips very gently and is rather evenly distributed as far as west as the eastern coast of Honshu and Hokkaido; thereafter, it deeps steeply to the north-west and is rather unevenly distributed, being highly clustered and reaching the depth of 500km in the southern part of the zone, but more dispersed and failing to penetrate the 410km discontinuity at the central and northern parts. Sub-crustal activity included thirty nine $M \geq 6.5$ events, fifteen of which occurred prior to the Tōhoku mega-event and seven were major (including the 2003/9/26, M8.3 Hokkaido event).

Figs. 8d and 9d respectively show the cumulative earthquake counts of crustal and sub-crustal seismicity from 2002.0 to 2011.19 (2011/3/10), just before the Tōhoku event. After that time, the catalogue is overwhelmed by the high volume and very long-lasting aftershock sequence, which extended both in and below the crust and completely obscured whatever background activity was there. For this reason, declustering has practically eliminated the post-Tōhoku seismicity and left so few "background" earthquakes, that little was to be gained by studying their cumulative count. As evident in Fig. 8d, crustal seismicity was marked by the extended aftershock sequences of the 2004/10/23 M6.8 and the 2008/6/14 M7.2 events, the rest contributing with relatively low-intensity and short-lived activity. Owing to their pronounced and easy to model character, foreground events have been effectively eliminated. Unfortunately this has left less than 100 events on which to base the analysis. The sub-crustal activity exhibits the expected quasi-linear increase decorated with time-local jerks due to low-intensity and short-lived aftershock sequences of major events (Fig. 9d, full catalogue). Interestingly enough, the 2003/9/16 M8.3 Hokkaido mega-thrust event had a rather minor aftershock signature, with other significant contributions coming from the 2003/5/26 M7.1 and 2005/11/15 M7.2 events; all of these have occurred at depths shallower than 45km.

## 4. RESULTS

The presence and intensity of correlation can be evaluated by means of the entropic indices $q_M$ and $q_T$ (Section 2). For obvious reasons, the appraisal of low-valued $q$ estimators can be ambiguous and for this reason, a rigorous threshold is required, on the basis of which to confidently decide whether a system is non-extensive or Poissonian. The determination of such *baselines* was taken up by Efstathiou et al. (2015; 2017) as well as Tzanis et al., (2018) who analysed a significant number of synthetic background catalogues generated on the basis of the ETAS model by the stochastic aftershock simulator "AFTsimulator" of Felzer (2007). This uses the Gutenberg-Richter and Omori-Utsu laws to simulate the statistical behaviour of background and foreground seismicity, and Monte Carlo methods to simulate background earthquakes as well as multiple generations of aftershocks, thus generating realistic synthetic catalogues consistent with the long-term seismotectonic characteristics of a given area (also see Felzer et al., 2002 and Felzer and Brodsky, 2006). The implementation was based on the ETAS parameterization obtained by declustering the catalogues of the South and North California Earthquake Data Centres, assuming uniform background seismicity rates with $b$=1 and a maximum expected magnitude of $M_L$=7.3.

With $\langle . \rangle$ denoting the expectation value, the analysis of many synthetic ETAS catalogues determined that $\max[\langle q_T(M_{th})\rangle + 3\sigma] \leq 1.15$ and that $\langle q_M(M_{th})\rangle$ exhibited almost *imperceptible* variation around 1.5, so that $b_q \approx 1$, consistently with the assumptions on which the catalogues were generated. Moreover, the entropic indices computed by grouping the synthetic earthquakes according to interevent distance yielded a $\max[\langle q_T(\Delta d)\rangle + 3\sigma] \leq 1.2$, while $\langle q_M(\Delta d)\rangle$ was also stable and exhibited small fluctuations around 1.5. It was thus safely concluded that experimental values of $q_T(M_{th}) \geq 1.15$ and $q_T(\Delta d) > 1.2$ would be compelling evidence of non-extensive dynamics. Last but not least, the baselines should be considered "absolute": the temporal entropic indices obtained from the synthetic catalogues deviate from the expectation value of unity because the AFTsimulator draws random numbers from a *uniform* distribution. On the other hand, experimental realizations of $q_T$ would be arbitrarily close to unity *only* for truly random (Gaussian) interevent time data. Conversely, the distribution of magnitudes in the synthetic catalogues was forced to follow the Gutenberg-Richter law and as a consequence, $q_M$ performs as expected. Although this issue was identified early on, we chose not to adjust the AFTsimulator code so as to establish rigorous criteria and to ensure that inference would be trustworthy and hard to contend. In order to facilitate the discussion, the following classification of the temporal entropic index will be used, according to which correlation is *insignificant* when $q_T$<1.15, *weak* when 1.15≤$q_T$<1.3, *moderate* when 1.35≤$q_T$<1.4, *significant* when 1.4≤$q_T$<1.5, *strong* when 1.5≤$q_T$<1.6 and *very strong* when 1.6≤$q_T$.

The analysis will entail:

**a)** Examination of the *long-term average state* of a seismogenetic system, i.e. its statistical nature over the entire period spanned by the catalogue. In full catalogues, the result would be dominated by the effect of local processes including, but not limited to, aftershock sequences. In the case of declustered catalogues the result would reflect mainly *global* processes distributed over the entire system. The analysis is performed by studying the variation of the entropic indices with respect to threshold magnitude ($M_{th}$) and interevent distance ($\Delta d$). Results are summarized in Table 2. In order to maintain experimental rigour, estimation of the entropic indices is *not* performed for catalogue subsets containing *less than* 500 events and results are *not* considered and displayed *unless* associated with a goodness of fit ($R^2$) *better* than 0.97.

**b)** Examination of the *evolution* of a seismogenetic system. In this case again, full catalogues include processes that are local in time *and* space such as swarms and aftershocks. Although in the context of SOC local processes can have global effects, the variation of $q$ with time will have to be influenced by localized activity for a period analogous to the duration of swarms and sequences. In declustered catalogues, the variation of $q$ with time is caused by *global* processes distributed over the entire seismogenetic system. Analysis is performed by forming F-M-T distributions over successive overlapping windows and solving them for the parameters of Eq. 7. Each window comprises 600 consecutive events and the overlap distance is 20-25 events. Unless otherwise specified, computations were performed using a threshold magnitude $M_{th}=3.0$. Again, for the sake of experimental rigour, we only consider results with goodness of fit *better* than 0.97. The discussion will focus on the values (level) and variation of the entropic indices. It is important to note that the entropic indices are computed as functions of the *natural time* of the seismogenetic process but are plotted as functions of calendar time. For a time series comprising $N$ events, the natural time of the occurrence of the $k^{th}$ event is the quantity $t_n=k/N$. (e.g. Varotsos et al., 2011). Given that the rate of event generation is variable and internally determined, this definition does away with the *scale* of the lapse between events while preserving their order and energy, concentrates on the dynamic expression of the causative process (seismogenesis). It follows that *corresponding* states of processes that are physically analogous but have different natural time characteristics, as for instance full vs. background earthquake occurrence, may not always appear to be synchronous when plotted against calendar time.

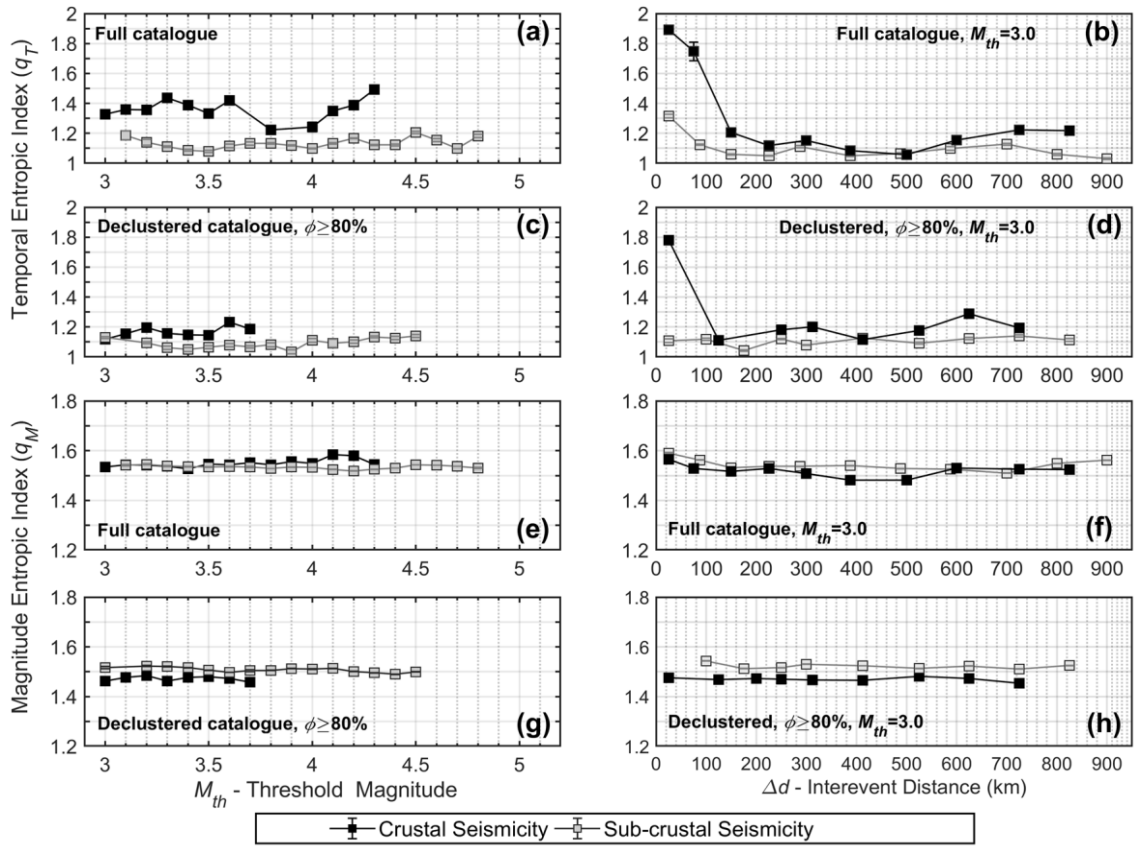

**Figure 10:** Analysis of the temporal (a-d) and magnitude (e-h) entropic indices determined for the full (a, b, e, f) and declustered (c, d, g, h) crustal and sub-crustal seismicity of the RKU source area. Results pertaining to *crustal* seismicity are illustrated with solid black squares and results pertaining to *sub-crustal* seismicity with solid grey squares. Panels (a, c) show the variation of the temporal entropic index ($q_T$) with threshold magnitude and panels (b, d) with interevent distance. Panels (e, g) show the variation of the magnitude entropic index ($q_M$) with threshold magnitude and panels (f, h) with interevent distance.

### 4.1 Ryukyu Arc

**a) Average State of Correlation in the Crust:** The variation of the full-catalogue temporal entropic index $q_T$ with threshold magnitude is shown in Fig. 10a, where it can be seen to oscillate between 1.22 and 1.44 (weak to significant correlation) and to have a mean value of 1.36±0.076 (moderate correlation). We are inclined to interpret the oscillatory nature of the entropic index as an effect of peculiarities in the data (magnitude re-porting) and not of the seismogenetic system. Beyond $M_{th}$=4.3 earthquake populations do not guarantee statistical rigour. On declustering the catalogue, $q_T$ drops to approximately 1.18 indicating a marginally correlated background (Fig. 10c and Table 2). This demonstrates that the moderate correlation observed in the full catalogue should be the result of aftershock and swarm activity. These the variation of the temporal entropic index with *interevent distance*, $q_T(\Delta d)$ is shown in Fig. 10b. When the analysis is carried out in such mode, one expects that $q_T(\Delta d)$ will have high values at $\Delta d$ shorter than 100km due to the dominant effect of short-range interaction in swarms and aftershock sequences. This is indeed observed in the full catalogue: $q_T(\Delta d \leq 100\text{km}) > 1.75$. For $\Delta d > 150$km the results indicate weak correlation: $q_T(\Delta d)$ is generally quite less than 1.25 and $\langle q_T(\Delta d > 150\text{km}) \rangle$= 1.15±0.062. At ranges longer than 600km $q_T$ rises to above 1.2 and given the elongate shape of RKU, this may indicate some degree of long-range interaction. On declustering the catalogue, one observes that $q_T > 1.75$ at $\Delta d \leq 100$km, but drops to under 1.15 at ranges of 100-200km only to increase to 1.2-1.3 at $\Delta d > 500$km (Fig. 10d). It follows that strong short-range and weak long-range correlation are persistent features of RKU crustal seismicity.

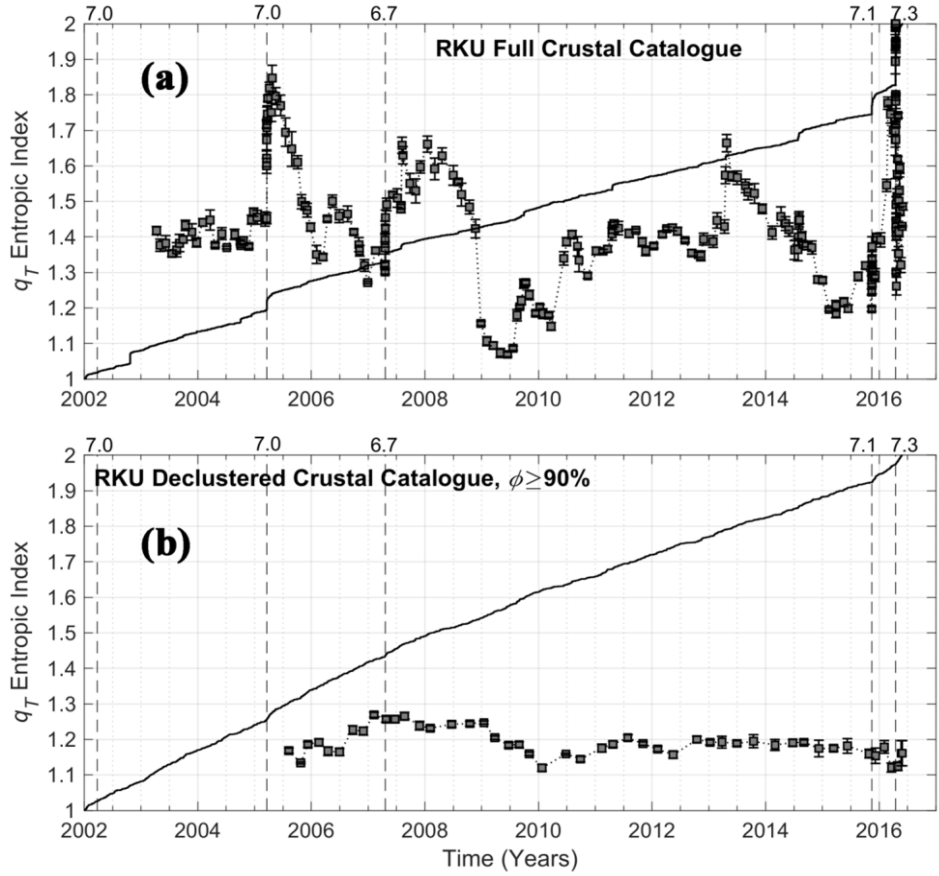

**Figure 11:** Evolution of the temporal entropic index during 2002-2016.5 for **(a)** the full and **(b)** the declustered *crustal* seismicity of the RKU source area. In both, the monotonically ascending solid lines represent the cumulative earthquake counts normalized and shifted so as to vary between 1 and 2. Broken vertical lines indicate the occurrence of $M \geq 6.7$ events.

The magnitude entropic indices $q_M(M_{th})$ and $q_M(\Delta d)$ do not exhibit variation worthy of attention, (Fig. 10e-h). However, as evident in Table 2, $q_M(M_{th})$ drops from a full catalogue mean value of 1.55±0.017, to declustered catalogue mean values of the order or 1.47±0.01; this corresponds to a change of mean $b_q$ values from 0.82±0.05 to 1.12±0.04. Likewise, $q_M(\Delta d)$ drops from a full catalogue mean of 1.52±0.025 to declustered catalogue means of the order of 1.47±0.01, again corresponding to mean $b_q$ changing from 0.93±0.09 to 1.13±0.04. The entropic index $q_M$, like the *b*-value to which it is related with, represents the scaling of the size distribution of earthquakes and is associated with the geometry of the active fault network and the homogeneity of the crust. Here it indicates a *sub-extensive* dynamic process exhibiting higher fault clustering in the full catalogue (lower *b* values) and lower in the declustered catalogues (higher *b* values).

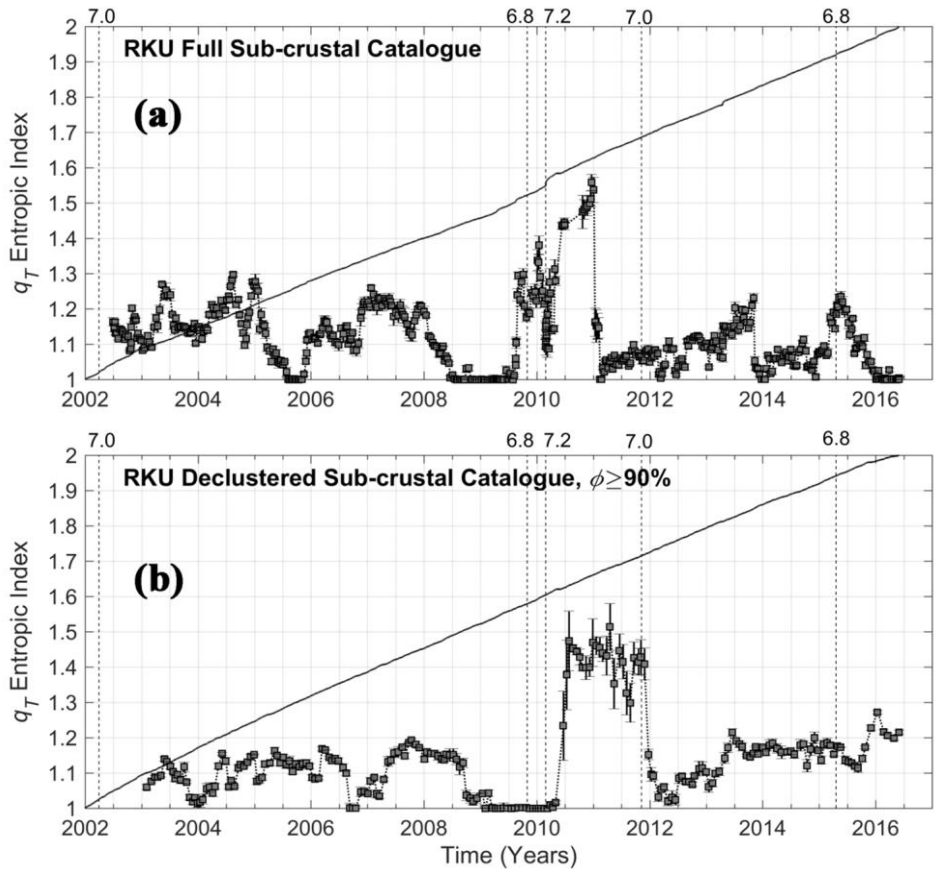

**Figure 12:** Evolution of the temporal entropic index during 2002-2016.5 for **(a)** the full and **(b)** the declustered *sub-crustal* seismicity of the RKU source area. In both, the monotonically ascending solid lines represent the cumulative earthquake counts normalized and shifted so as to vary between 1 and 2. Broken vertical lines indicate the occurrence of $M \geq 6.8$ events.

**b) Average State of Correlation below the Crust:** In Fig. 10a, it is straightforward to see that the full-catalogue $q_T(M_{th})$ is almost featureless and consistently lower that 1.2, averaging to $1.13 \pm 0.035$ (Table 3) and indicating a dominantly Poissonian seismogenetic process. On declustering the catalogue, $q_T(M_{th})$ remains featureless and drops even more to an averages value of 1.08-1.09 (Fig. 10c and Table 3). The variation of the sub-crustal temporal entropic index with interevent distance is again shown in Fig. 10b. For $\Delta d < 50$km, the full catalogue yields a $q_T$ value of 1.32, reflecting short-range interaction in the low-intensity and short-lived sub-crustal aftershock sequences and other dependent events. For $\Delta d > 100$km, the full-catalogue $q_T$ is featureless and mostly lower than 1.10, averaging to $1.07 \pm 0.03$. On declustering the catalogue, the weak the short-range correlation *disappears* and the still featureless $q_T(\Delta d)$ remains consistently lower than 1.15, consistently with the full catalogue. All in all, the dynamics of the sub-crustal active fault network in RKU apprar to be *practically* Poissonian.

As with crustal seismicity, the magnitude entropic indices $q_M(M_{th})$ and $q_M(\Delta d)$ do not have noteworthy characteristics (Fig. 10e-h). The full catalogue $q_M(M_{th})$ averages $1.53 \pm 0.01$ and the declustered catalogue $q_M(M_{th})$ to $1.51 \pm 0.01$, meaning that they are comparable. Likewise, the full catalogue $q_M(\Delta d)$ averages to $1.54 \pm 0.01$ and the declustered catalogue $q_M(\Delta d)$ to $1.52 \pm 0.01$. This very small change is a consequence of the low-intensity aftershock activity and has minimal effect on the distribution of active faults at local and regional scales.

**c) Evolution of Correlation in the Crust:** The evolution of the temporal entropic index is illustrated in Fig. 11a for the full catalogue and Fig. 11b for the version declustered at the 90% probability level. The full catalogue $q_T(t)$ exhibits frequent dynamic variation. As expected, very strong correlation ($q_T > 1.8$) is observed in association with aftershock sequences of major earthquakes and specifically with that of the M7 event of 2005/3/20, the M6.7 event of 2007/4/20 and the M7.1 and M7.3 events of 2015/11/14 and 2016/4/16. Correlated states associated with aftershock sequences decay fairly rapidly. In periods without major events, the evolution of $q_T(t)$ is also dynamic and exhibits both fast and slow variation. The former can almost always be seen to associate with jerks of the cumulative event count, typically attributable to localized activity. The rapid succession of such effects also appears to combine and keep $q_T(t)$ at levels significantly higher than 1.3, for most of the time. Examples of slow variation are the strong (1.5–1.65) correlation observed between years 2007 and 2009, or the upward trend observed between years 2010 and 2013. In the declustered catalogue the variation of $q_T(t)$ is minimal: it is almost flat and definitely above the randomness threshold of 1.15, exhibiting values consistent with those of Table 2. This appears to confirm the conclusion reached above, that weak long-range interaction is a persistent global feature of RKU crustal seismicity. Notably, during 2007–2009, $q_T(t)$ is on average higher than 1.25, whereas it is slightly lower than 1.2 for the rest of the time. The effect is significant at the 95% confidence level and corresponds to the period of strong correlation observed in the full catalogue (Fig. 11a), meaning that its origin, whatever of nature, was a global-scale process and possibly associated with the significant events of 2003-2007.

**d) Evolution of Correlation below the Crust:**
The full-catalogue $q_T(t)$ is shown in Fig. 12a and exhibits significant short-term variability that is always limited to levels lower than 1.3 (weak correlation) and at times drops to 1-1.05 (randomness). Only exception is in the period 2009.6–2011.9 in which three significant events have taken place: on 2009/10/30 with M6.8, on 2010/2/27 with M7.2 and on 2011/11/8 with M7. In that period the correlation increased and varied dynamically, becoming strong (>1.5) for a short time (2010.8–2011). Notably, after the M7 event of 2011/11/8, $q_T(t)$ dropped to levels lower than 1.2. The $q_T(t)$ curve obtained from the catalogue declustered at the 90% probability level is shown in Fig. 12b: it is overall smoother than its full catalogue counterpart and overall stabilized at a level of 1.1–1.2, consistently with the results shown in Fig. 10. Again, exception is the period 2010.8–2011, in which $q_T(t)$ rises to moderate levels (1.4–1.5). As the 90% declustered catalogue represents global processes, this result demonstrates that subcrustal regional fault networks may experience transitions between uncorrelated and correlated states, in response to (unspecified) causative factors that act on the fault network(s) over long ranges and possibly include the production of large earthquakes. This point will be revisited in Section 5.

### 4.2 Philippine Sea – Pacific Convergence

**a) Average State of Correlation in the Crust:**
The full-catalogue variation of the temporal entropic index with magnitude is shown in Fig. 13a. Up to $M_{th}=4$, the index varies between 1.2 and 1.3 and then drops slightly to approx. 1.15: overall weak correlation is more or less evenly distributed over all magnitude scales. On declustering the catalogue, $q_T(M_{th})$ drops to 1.1–1.2 indicating a background verging on randomness (Fig. 13c and Table 2). The full-catalogue variation of $q_T$ with interevent distance is shown in Fig. 13b. As expected, it is very high (1.96) at ranges $\Delta d \leq 100$km, mainly due to the dominant effect of the extended aftershock sequence of the M7.4 event of 2010/12/22, as well as of the low-intensity, shorter-lived sequences of the M6.5 event of 2005/2/10, and the M6.8 event of 2006/10/24. At ranges between 100 and 350 km, $q_T(\Delta d)$ varies between 1.1 and 1.21, indicating insignificant to rather weak correlation. At the same time it exhibits a clear upward trend so as to increase to levels significantly above 1.2 at $\Delta d > 400$km (weak to moderate): the level of $q_T$ values observed at such ranges are clearly indicative of operational long-range interaction. On declustering the catalogue short-range correlation cannot be evaluated, but the upward trend persists so that $q_T$ rises from approximately 1.2 at $\Delta d=100$km to approximately 1.3 at $\Delta d > 600$km, thus confirming the presence of weak to moderate long-range interaction (Fig. 13d). As with RKU, weak to moderate long-range correlation is a persistent global feature of PSP crustal seismicity.

As shown in Fig. 13e-f, the full-catalogue magnitude entropic indices $q_M(M_{th})$ and $q_M(\Delta d)$ are stable and featureless with $\langle q_M(M_{th}) \rangle = 1.56 \pm 0.01$ and $\langle q_M(\Delta d) \rangle = 1.59 \pm 0.02$ indicating a rather high degree of fault clustering. The declustered catalogues exhibit a slight, quasi-linear decrease of $q_M$ with magnitude. At the 80% probability level shown in Fig. 13b, $q_M(M_{th})$ drops from $1.56 \pm 0.0014$ at $M_{th}=3$ ($b_q=0.77$) to $1.51 \pm 0.0023$ at $M_{th}=4$ ($b_q=0.96$). Note also that the declustered-catalogue $q_M(\Delta d)$ changes from 1.55 at $\Delta d=100$km to 1.6 at $\Delta d > 200$km (Fig. 13f). These results may imply that on a regional level, smaller active faults are relatively fewer and maybe tightly clustered around larger faults that are broadly distributed and quasi-independent.

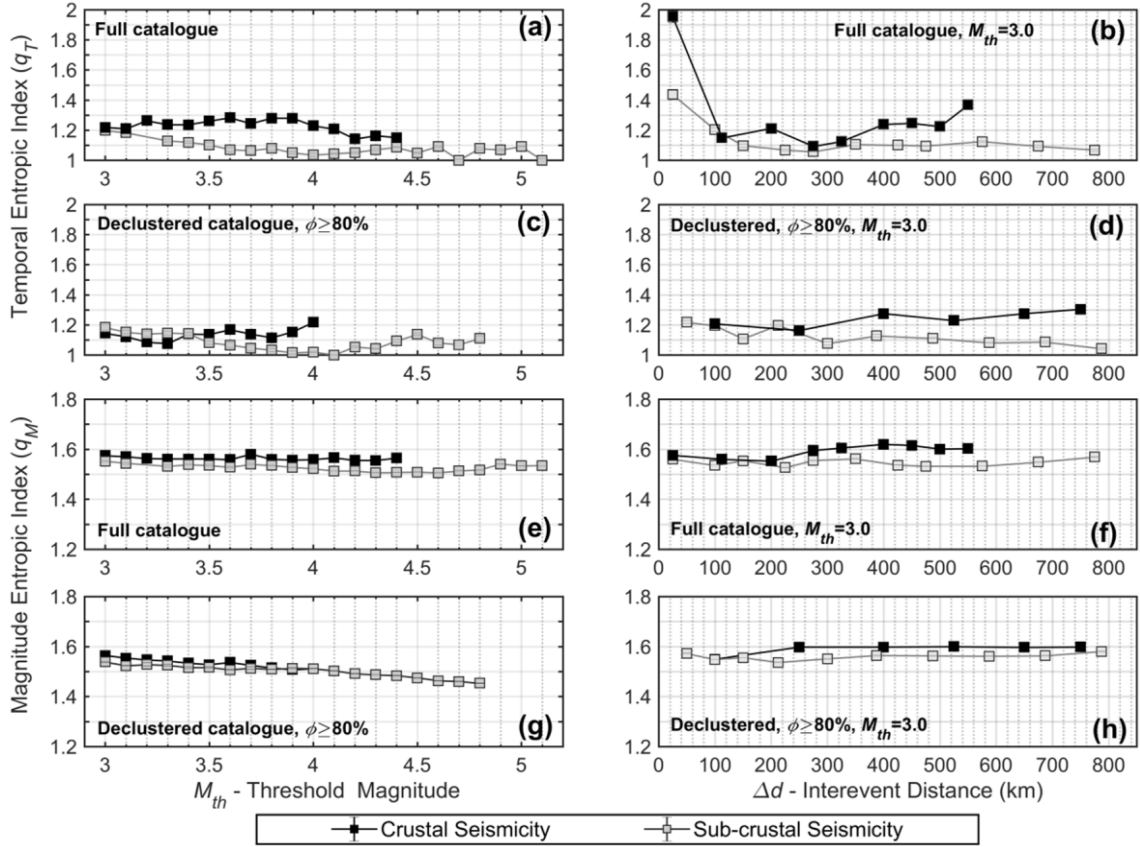

**Figure 13:** As per Fig. 10 but for the PSP source area.

**b) Average State of Correlation below the Crust:** Fig. 13a clearly illustrates that the full-catalogue sub-crustal temporal entropic index steadily decreases from a high value of 1.22 at $M_{th}=3$ to under 1.1 at $M_{th} \geq 4$; together with an average value of 1.08±0.049 (Table 3), this demonstrates a practically random seismogenetic process. The same conclusions are drawn by studying declustered versions of the sub-crustal catalogue (Fig. 13c and Table 3). The full-catalogue sub-crustal $q_T(\Delta d)$ is shown in Fig. 13b. Remarkably, at interevent distances shorter than 100km $q_T$ is only 1.44±0.02, indicating moderately correlated short-range processes, possibly due to the relatively short duration and low intensity of aftershock activity in combination with the high productivity of small events everywhere in the subduction zone. At longer interevent distances, $q_T$ drops to under 1.12. On declustering the catalogue, the moderate short range correlation disappears and $q_T(\Delta d)$ remains consistently lower than 1.15, gradually diminishing with distance and averaging to 1.12±0.06 (Fig. 13d). All in all, sub-crustal seismogenesis is practically Poissonian.

The full-catalogue magnitude entropic indices are stable, featureless and comparable (Fig. 13e-f), with $\langle q_M(M_{th}) \rangle = 1.53 \pm 0.01$ and $\langle q_M(\Delta d) \rangle = 1.55 \pm 0.01$; this indicates a very similar and not particularly high degree of active fault clustering. The declustered-catalogue $q_M(M_{th})$ also exhibits quasi-linear decrease with magnitude (Fig. 13g): at the 80% probability level, it declines from 1.54 for $M_{th}=3$ ($b_q=0.85$) to 1.45 for $M_{th}=4.8$ ($b_q=1.21$). In Fig. 13f, the declustered-catalogue $q_M(\Delta d)$ does not exhibit variation and fluctuates about a mean value of 1.56±0.01, which is congruent with the corresponding result of the full catalogue. As in the case of background crustal seismicity, these results are consistent with small earthquake activity taking place in short-lived clusters associated with quasi-independent larger active faults. Given the quasi-Poissonian nature of sub-crustal seismogenesis, larger faults appear to be uncorrelated (non-interacting) in the Wadati-Benioff zone.

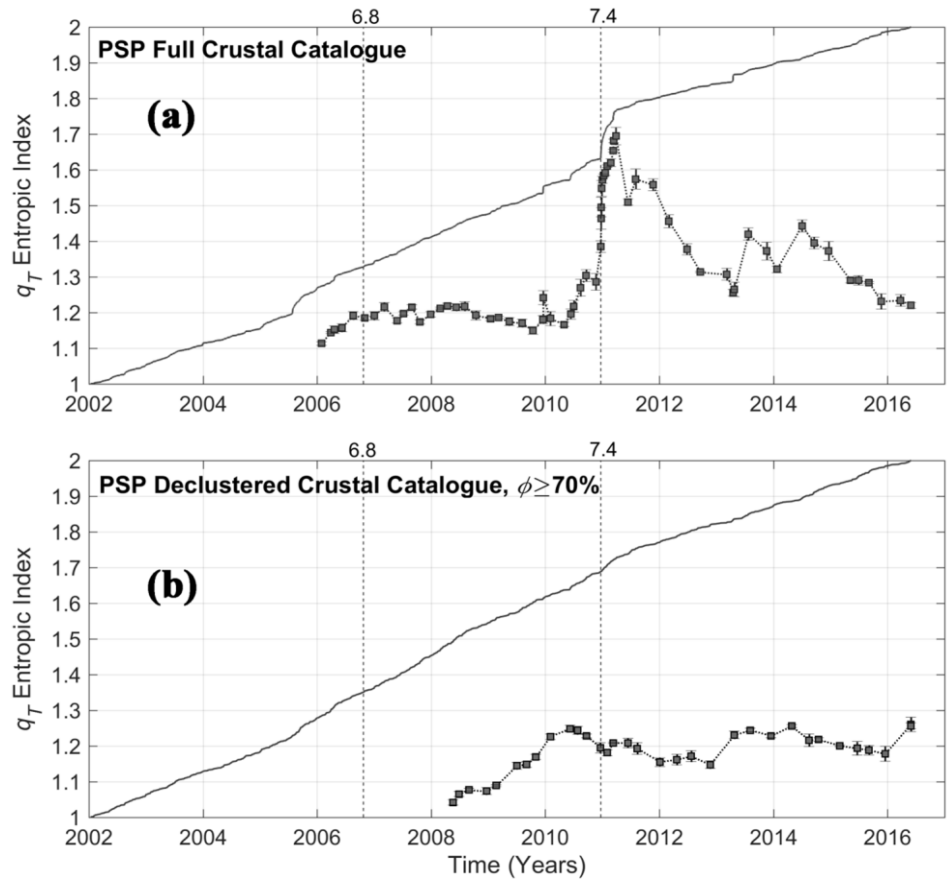

**Figure 14:** Evolution of the temporal entropic index during 2002-2016.5, for **(a)** the full and **(b)** the declustered *crustal* seismicity of the PSP source area. The monotonically ascending solid lines represent the cumulative earthquake counts normalized and shifted as to vary between 1 and 2. Broken vertical lines indicate the occurrence of *M*≥6.7 events.

**c) Evolution of Correlation in the Crust:** The variation of $q_T$ with time is illustrated in Fig. 14a for the full crustal catalogue and Fig. 14b for the version declustered at the 70% probability level; at higher probabilities the scarcity of "background" events does not allow for meaningful results. It is evident that up to the occurrence of the M7.4 event of 2010/12/22, the full-catalogue index indicates a weakly correlated process with $q_T(t)\approx1.2$, although its variation is decorated with local excursions corresponding to short-lived sequences of intermediate-sized events. From that time on, and up to mid-2011, correlation becomes strong to very strong ($q_T>1.5$) due to the aftershock sequence of that event and subsequently relaxes to moderate-significant ($q_T \sim 1.3–1.5$). It should also be noted that although the aftershock sequence of the 2010/12/22event died out after 2011.5, the correlation remained significant across the PSP crust and diminished slowly. Analogous observations can be made in the background process (Fig. 14b): $q_T(t)$ was less than 1.1 in 2008.4, but rose to above 1.2 by early 2010 and remained at levels indicating weak but *persistent global* correlation. It thus appears that the PSP crustal fault network developed a degree of long-range interaction as of the beginning to year 2010.

**d) Evolution of Correlation below the Crust:** The variation of $q_T$ with time is illustrated in Fig. 15a for the full sub-crustal catalogue and Fig. 15b for the version declustered at the 90% probability level. Observations analogous to those made for RKU sub-crustal seismicity can also be made here. The full-catalogue $q_T(t)$ exhibits substantial variability but is generally limited to levels lower than 1.2 and frequently drops to 1-1.05 (randomness). There are two exceptions. The first is in the period 2005–2006.5 and includes the 2004/5/30 M6.7 and 2005/1/19 M6.8 events. With hypocentral depths of 23 and 31 km respectively, these appear to have generated "normal" aftershock sequences, at the peak of which $q_T$ is higher than 1.5 (strong correlation). The second is in the period 2010–2012.6 which is characterized by a series of large-major earthquakes, namely the M6.8 event of 2009/8/9, the M7.1 event of 2010/11/30 and the M7 event of 2012/1/1); these have occurred at depths of 300 to 500 km but also appear to be associated with "normal" aftershock sequences. During this period $q_T(t)$ varies between 1.5 and 1.7 (strong to very strong correlation) and is particularly oscillatory. The deep major event of 2015/5/30 ($M = 8.1$, $D \approx 680$km) has insignificant aftershock sequence and no imprint on the evolution of $q_T$. In periods without major events, $q_T(t)$ exhibits short-term variability that in most cases coincides with sequences of intermediate-sized events. The declustered catalogue (Fig. 15b) yields a $q_T(t)$ function generally lower than 1.15 (insignificant correlation) and much smoother than that of the full catalogue. Exception, again, is the period 2010–2013 during which $q_T(t)$ rises to levels of 1.3 to 1.5 (moderate to significant correlation). Presumably, this occurs in association with the unrest that generated the three earthquakes. One may also point out that the onset of this period coincides with the onset of the period of weak correlation in the crust (Fig. 14b).

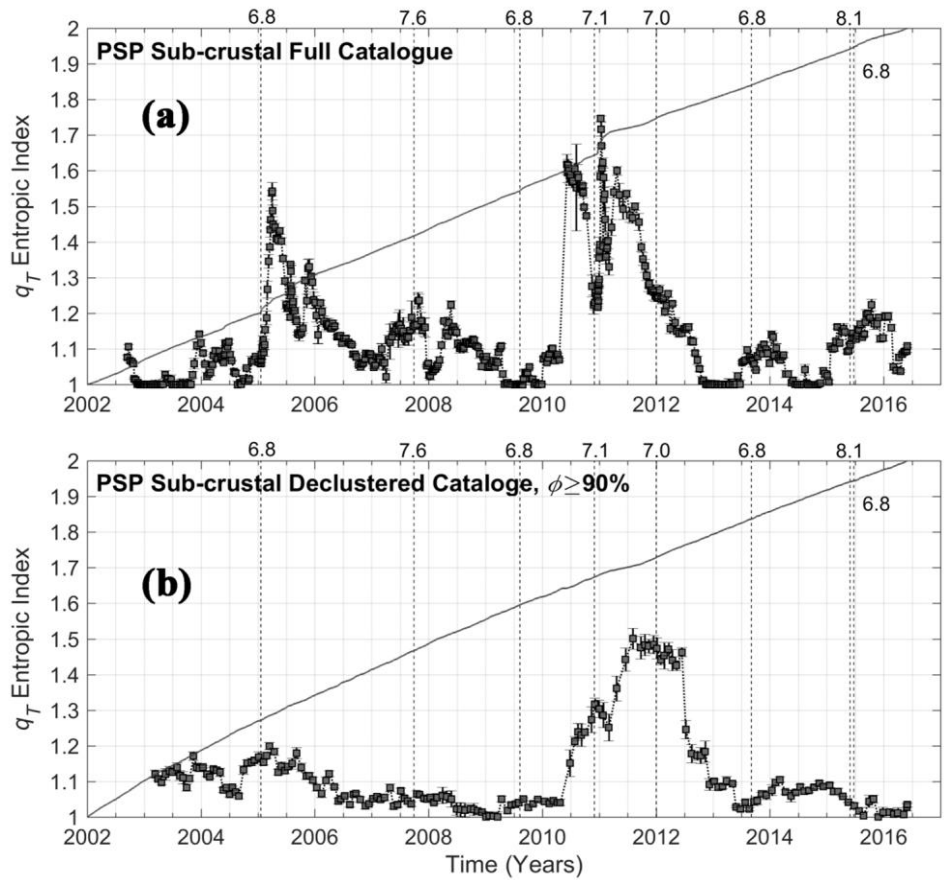

**Figure 15:** Evolution of the temporal entropic index during 2002-2016.5 for **(a)** the full and **(b)** the declustered *sub-crustal* seismicity of the PSP area. Monotonically increasing solid lines represent the cumulative earthquake counts normalized and shifted so as to vary between 1 and 2. Broken vertical lines indicate the occurrence of $M \geq 6.8$ events.

Given that only global processes should remain in a catalogue declustered at the 90% probability level, this demonstrates that crustal and sub-crustal regional fault networks may transition between uncorrelated and correlated states in response to (geodynamic?) factors that organize (?) them over long ranges and generate significant earthquakes.

### 4.3 South-West Japan

**a) Average State of Correlation in the Crust:** As seen in Fig. 16a, the full-catalogue temporal entropic index increases from 1.18 at $M_{th}=3$ to 1.51 at $M_{th}=3.7$ exhibiting persistent quasi-linear trend. Fig. 16c illustrates the same for the catalogue declustered at the 70% probability level. The level of correlation is similar to that of the full catalogue and the linear upward trend is still present: $q_T$ varies from 1.09 at $M_{th}=3$ to 1.34 at $M_{th}=3.4$, increasing with a slope of approximately 0.65 per magnitude unit. Declustering at higher probability levels yields almost identical results. The variation of the full-catalogue $q_T$ with interevent distance is shown in Fig. 16b. It is apparent that for $\Delta d \leq 100$km, $q_T(\Delta d) \approx 1.9$ indicating very strong short-range correlation, which is partly attributable to the extended aftershock sequence of the M6.9 events of 2007/3/25. It is also apparent that for ranges beyond 200km the results indicate rather very weak correlation: $1.15 \leq q_T(\Delta d) \leq 1.2$. Turning to the declustered catalogue one observes that for $\Delta d \leq 100$km, $q_T(\Delta d) > 1.3$ while for $\Delta d > 250$km $q_T$ drops to under 1.15 (Fig. 16d). This demonstrates absence of long-range interaction. On considering all the evidence of Fig. 16a-d, one may conclude that the increase of $q_T$ with magnitude observed in Fig. 16a is an effect involving progressive increase of interaction between progressively larger events occurring at short and intermediate ranges.

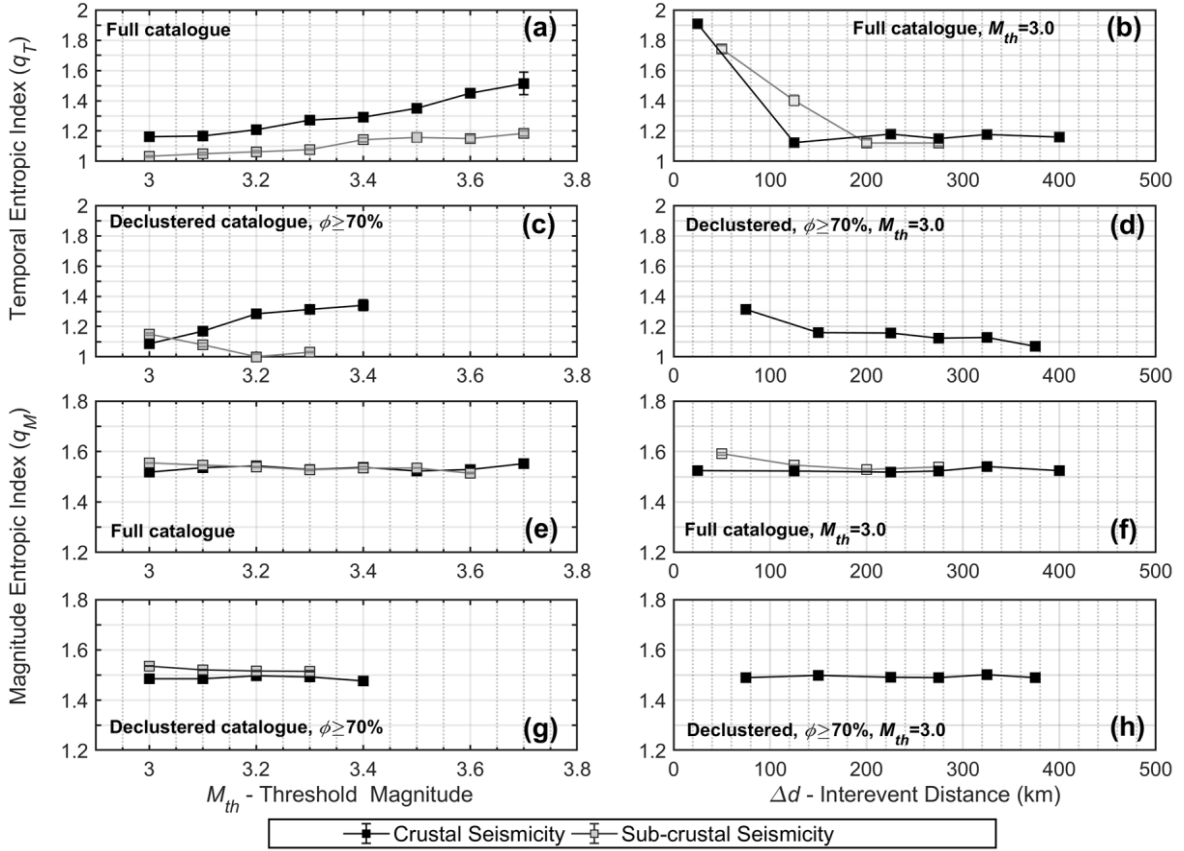

**Figure 16:** As per Fig. 10 but for the SJP source area.

The magnitude entropic index does not have noteworthy characteristics both as function of $M_{th}$ and $\Delta d$. The full-catalogue $q_M(M_{th})$ fluctuates slightly about a mean of $1.53 \pm 0.01$ and the declustered-catalogue equivalents about a mean of $1.49 \pm 0.01$ (Table 2, Fig. 16e and 16g). The small reduction of 0.04 can easily be explained as an effect of declustering: it corresponds to a $b_q$ value change from $0.9 \pm 0.03$ to $1.05 \pm 0.03$, i.e. from higher to lower clustering of fault activity. Similarly, the full-catalogue $q_M(\Delta d)$ fluctuates slightly about $1.53 \pm 0.01$ and its declustered-catalogue equivalents about $1.49 \pm 0.005$, (Fig. 16f and 16h respectively), and the corresponding small reduction admits the same interpretation. These results indicate that within the period of observations, the distribution and scaling of active faulting is more or less uniform across SJP.

**b) Average State of Correlation below the Crust:** Fig. 16a shows that the full-catalogue temporal entropic index increases with magnitude, from 1.03 at $M_{th}=3$ to 1.18 at $M_{th}=3.7$. As far as the declustered catalogues are concerned, results could be obtained only at the 70% probability level and only for a limited magnitude range (Fig. 16c); with $q_T(M_{th}) \leq 1.15$ throughout, it is evident that background sub-crustal seismicity comprises randomly occurring events. Fig. 16b illustrates the variation of the full-catalogue temporal entropic index with interevent distance. For $\Delta d < 200$km $q_T$ is 1.74–1.4 but for $\Delta d \geq 200$km it drops to the level of 1.1. For the most part, this short to intermediate range correlation, is conveniently explainable by the extended aftershock sequence of the M7.1 and M7.4 events of 2004/9/5. It was not possible to obtain statistically significant results for any declustered realization of the sub-crustal catalogue. The limited evidence affordable by the declustered SJP catalogue does not allow insight into the background seismogenetic process, although it points to it being practically Poissonian.

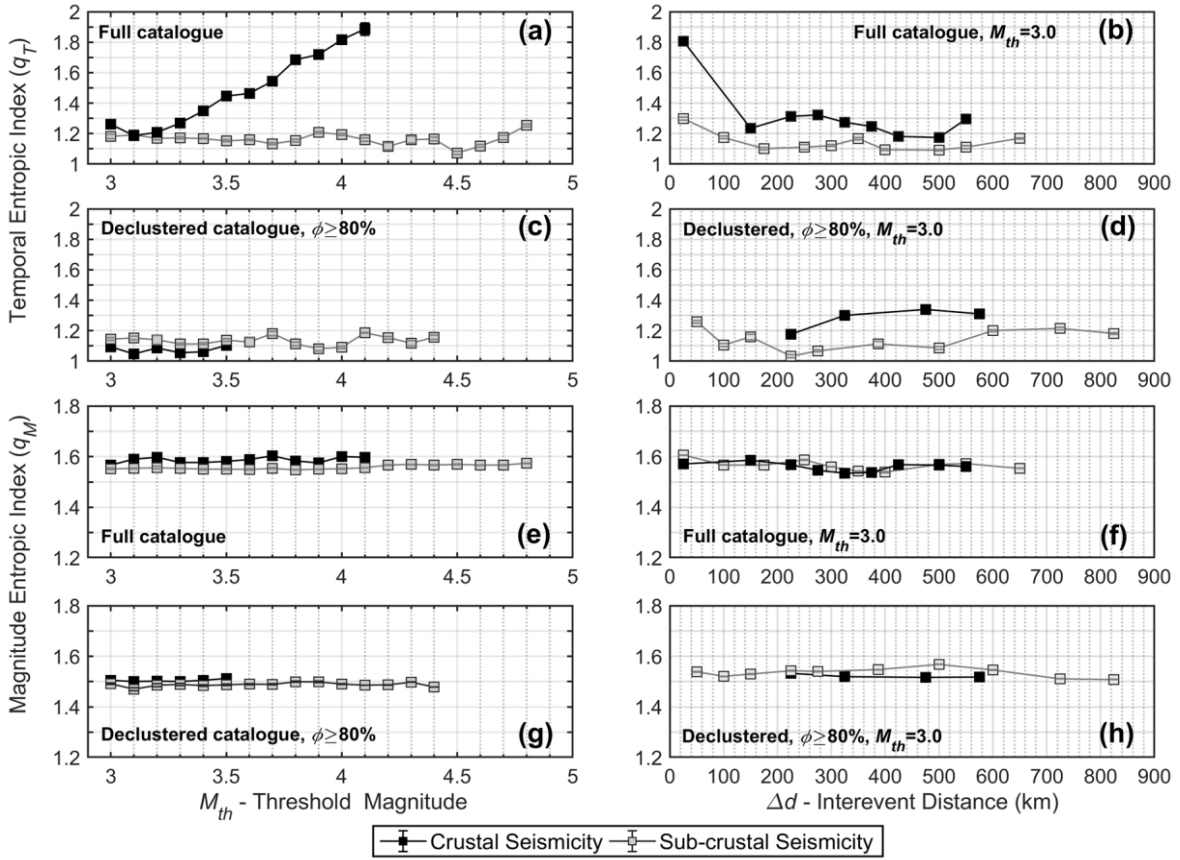

**Figure 17:** As per Fig. 10 but for the OKH source area and for the period 2002-2011.19.

The magnitude entropic index does not show features worthy of attention, both as function of $M_{th}$ and $\Delta d$. The full-catalogue $q_M(M_{th})$ fluctuates slightly about 1.54±0.01 and its declustered counterpart about 1.52±0.01 (Table 3 and Fig. 16e,g); the difference is very small to be significant. As evident in Fig. 16f, the full-catalogue $q_M(\Delta d)$ is approximately 1.59 for $\Delta d$<100km and drops to approximately 1.54 thereafter. Although it is rather difficult to provide rigorous explanation, the increased short-range clustering implied by these results could be attributed to aftershock sequences.

### 4.4 The Okhotsk Plate System and the Pacific – Okhotsk Convergence

This broad area is particularly interesting because of the 2011/3/11 M9 Tōhoku mega-thrust event.

However, detailed analysis has to be limited to the period 2002/1/1 - 2011/3/10 (2011.19). After that date, earthquake activity essentially comprises a colossal and very strongly correlated aftershock sequence: $q_T(M_{th})$>1.8 throughout. Because of this, declustering leaves fewer than 1000 "background" events both in and below the crust – very few to work with. Finally, the rupture was located near the base of the crust and the aftershock sequence developed so as to not warrant separation of "crustal" and "sub-crustal" events on a depth-of-Moho basis.

**a) Average State of Correlation in the Crust:** The full-catalogue temporal entropic index vs. threshold magnitude is shown in Fig. 17a; at small magnitudes it indicates weak sub-extensivity, but for $M_{th}$≥3.3 increases steeply and quasi-linearly so as to attain a level of 1.89 by $M_{th}$=4.1 (very strong sub-extensivity). On declustering the catalogue, $q_T$ drops dramatically to lower than 1.1, indicating a practically random background (Fig. 17c and Table 2). The variation of the temporal entropic index with interevent distance is shown in Fig. 17b and 17d. In the full catalogue and for $\Delta d$<100km, $q_T$=1.81 and indicates very strong short-range cor-relation, as expected. At longer interevent distances $q_T$ indicates weak to moderate correlation with particular reference to the interval 200–400km. On declustering the catalogue one is surprised to see that $q_T$ cannot be computed for ranges shorter than 200km (Fig. 17d) due to the small number of "background" earthquakes (235 to be exact); for $\Delta d$>200km $q_T$ varies between 1.18 and 1.34 (moderate long-range correlation) consistently with the results of the full catalogue. All in all, it can ve safely concluded that the overall strong correlation observed in Fig. 17a was a result of intense short-range interaction within clusters and that moderate long-range interaction is a *persistent* global feature of crustal seismicity in the Okhotsk plate.

The magnitude entropic indices $q_M(M_{th})$ and $q_M(\Delta d)$ are very stable and consistent (Fig. 17e-h). However, it is worth noting that $q_M(M_{th})$ drops from a full-catalogue mean value of 1.58±0.017, to declustered-catalogue mean values of the order of 1.50±0.01 (Table 2); this corresponds to a change in $b_q$ from 0.72±0.05 to 0.98±0.02. Likewise, $q_M(\Delta d)$ drops from the full-catalogue mean value of 1.56±0.017 to declustered-catalogue mean values of the order of 1.52±0.01; this again corresponds to a change in $b_q$ from 0.79±0.05 to 0.92±0.03. The results are significant at the 95% level and indicate relaxation in the clustering of earthquake activity due to the removal of aftershocks.

**b) Average State of Correlation below the Crust:** Sub-crustal seismic activity is mainly due to the Wadati-Benioff zone of the Pacific–Okhotsk convergence. In Fig. 17a, the full-catalogue temporal entropic index is an almost featureless function of magnitude and indicates a borderline seismogenetic process: $q_T$ has a mean value of 1.16±0.04. On declustering the catalogue, $q_T(M_{th})$ remains featureless and fluctuates about a mean of 1.12 at all probability levels (Fig. 17c and Table 3). It can safely be concluded that correlation is *insignificant* in both the foreground and background processes. With respect to interevent distance, the full catalogue yields a $q_T$ of 1.3 at ranges $\Delta d$<50km presumably due to moderate correlation in the short-lived sub-crustal aftershock sequences(Fig. 17b); subsequently the index drops and varies between 1.09 and 1.17 indicating practical absence of long-range interaction. On declustering the catalogue, only weak correlation persists at short ranges (Fig. 17d); at intermediate and long-ranges $q_T(\Delta d)$ remains consistently lower than 1.2 and averages to 1.13±0.06. It appears that the sub-crustal fault network has not yet decided if it is Poissonian of marginally sub-extensive (at least within the period of observations).

The sub-crustal magnitude entropic indices $q_M(M_{th})$ and $q_M(\Delta d)$ do not possess notable characteristics (Fig. 17e-h). As evident in Fig. 17e,g and Table 3, all realizations of $q_M(M_{th})$ are very consistently determined but the full-catalogue realization averages to 1.56±0.08 and the declustered catalogue versions average to 1.49–1.50; this would again seem to imply relaxation due to removal of dependent events. The experimental determinations of $q_M(\Delta d)$ point to, but do not verify this inference because they are not as stable: the full-catalogue $q_M(\Delta d)$ averages to 1.57±0.02, while all the declustered versions to 1.54±0.018.

**c) Evolution of Correlation:** The variation of $q_T$ with time is illustrated in Fig. 18a for the full and declustered ($\phi$≥80%) crustal catalogue, in Fig. 18b for the full sub-crustal catalogue and in Fig. 18c for the declustered sub-crustal catalogue ($\phi$≥80%). In the full crustal catalogue, one may observe states of very strong correlation ($q_T$>1.8) immediately after large events, which gradually relax to states of reduced correlation as aftershock sequences die away. The same pattern may be observed in the full sub-crustal catalogue, albeit with two noticeable differences. First, the intensity (level) or correlated states is significantly lower: with the exception of the M8.3 Hokkaido event of 26/9/2003, the $q_T$ observed immediately after large events is generally under 1.5 (moderate to strong). Second, the decay of correlated states is much faster and the relaxed states are practically Poissonian ($q_T$ is generally lower than 1.15). Once again, when catalogues with aftershock sequences are considered, crustal fault networks turn out to be significantly more correlated than sub-crustal networks and slower to relax.

Two possibly significant observations are the following: **a)** During 2009–2011.19 and immediately following the aftershock sequence of the shallow M7.2 event of 14/6/2008, the intensity of correlation in the crust *did not* die away; $q_T$ remained at the level of 1.5 to 1.6 (Fig. 18a). **b)** For a duration of approximately 6 months between epochs 2010.44 and 2010.97, the intensity of sub-crustal correlation jumped from under 1.2 to 1.4–1.5 (significant to strong). These effects may indicate that the crustal and sub-crustal fault networks may have been locked into states of elevated correlation, pos-sibly as part of the process leading to the M9.3 Tōhoku mega-event; the possibility appears to be corroborated by the analysis of the declustered crustal and sub-crustal catalogues.

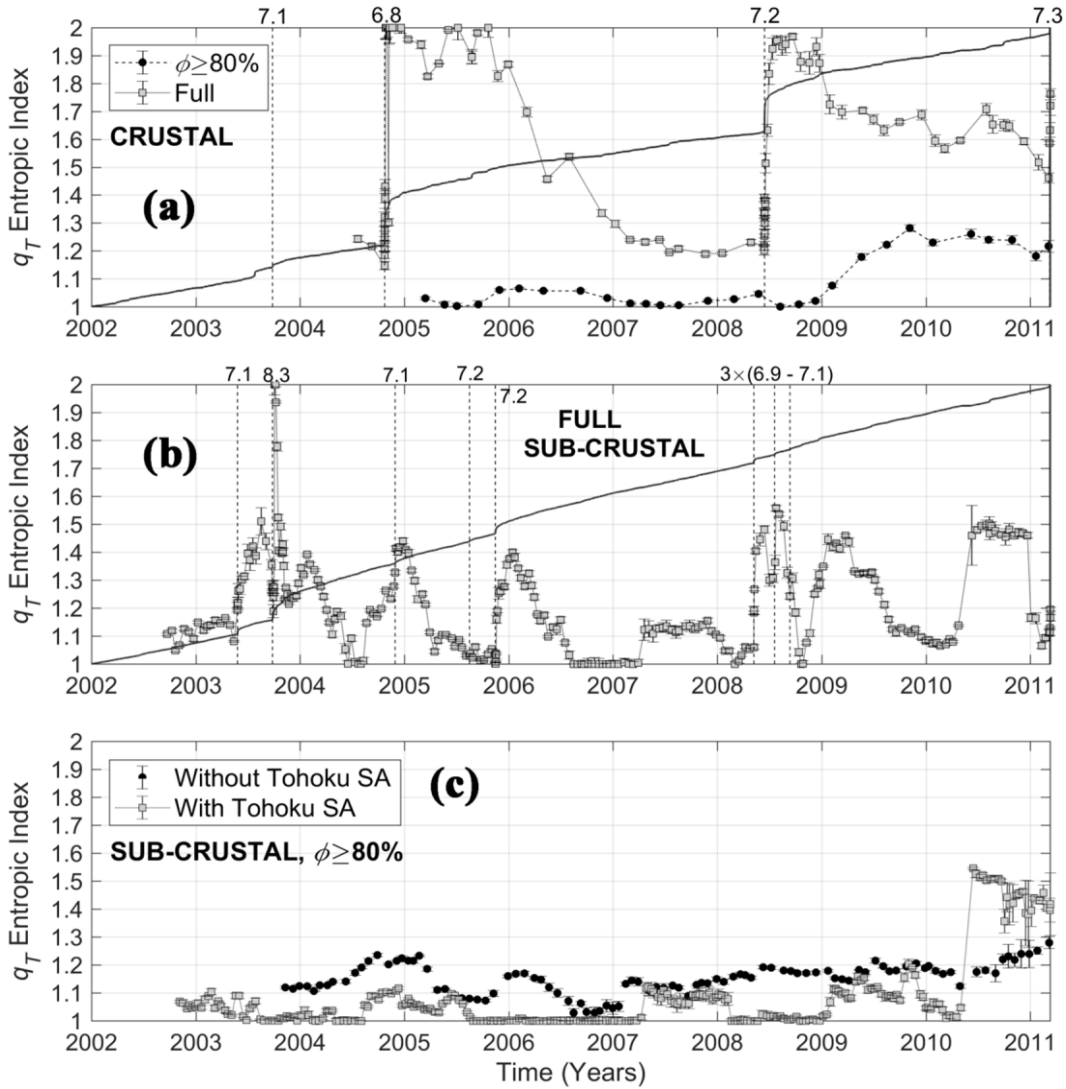

**Figure 18:** Evolution of the temporal entropic index in OKH during the period 2002-2011.19. **(a)** Crustal full (solid grey squares) and declustered (black solid circles) seismicity. **(b)** Sub-crustal *full* seismicity. **(c)** Sub-crustal *declustered* seismicity over the entire OKH (grey squares) and by *excluding* the source area of the 11/3/2011 M9.3 Tōhoku mega-earthquake (black solid circles). In (a) and (b) the monotonically ascending solid lines represent the respective cumulative earthquake counts normalized and shifted so as to vary between 1 and 2. Broken vertical lines indicate the occurrence of *M*≥6.9 events.

In Fig. 18a, the declustered-catalogue crustal $q_T(t)$ is illustrated with solid black circles. A caveat here is that due to very small number of background events the analysis was carried out with a sliding window size of 400 events (395 degrees of freedom), as opposed to the 600 events (595 degrees of freedom) used everywhere else: $q_T(t)$ estimators, albeit stable and consistent, are respectively less rigorous from a statistical point of view. At any rate, the results indicate that up to year 2009.0 the *crustal* background process in OKH was practical- ly Poissonian ($q_T<1.1$), and that after 2009.8 it gradually developed to *weak correlated*. In Fig. 18c, light grey squares represent the evolution of the temporal entropic index for the declustered catalogue of the *entire* study area and solid black circles the same for a subset in which we have *excluded* the source area of the Tōhoku event (as defined by the distribution of aftershocks). It is evident that up to year 2010, the intensity of correlation is generally low but the temporal entropic index estimators are *markedly* higher when the Tōhoku Source Area (SA) is excluded: the mean value of $q_T$ *without* Tōhoku was 1.14±0.051 while the mean value of $q_T$ *with* Tōhoku was only 1.04±0.046. This indicates that the sub-crustal background seismicity of the Tōhoku source area was purely Poissonian up to year 2010, and its admixture with the weakly correlated background of the *rest* of the subduction zone resulted in a randomized catalogue. More importantly however, during the epoch 2010.3-2011.19 the evolution of $q_T$ *excluding* the Tōhoku SA exhibited a steady increase from 1.13 to 1.28 (upper echelons of weak correlation), while the evolution of $q_T$ *including* Tōhoku jumped from under 1.1 to 1.4-1.55, (insignificant to *strong*). It can thus be inferred that shortly prior to the Tōhoku event, the entire Pacific–Okhotsk subduction entered a period of increased non-equilibrium and that the Tōhoku SA in particular, switched to an intensely non-equilibrating state in which it remained until the occurrence of the mega-earthquake. These results should be viewed with prudence. The validity of experimental results is commensurate to the quality of the data and while have no reason to question the data (catalogue), our capacity to offer explanation as to why the Tōhoku SA behaved in such way is hampered by the complete absence of analogous observations from analogous seismogenetic areas.

## 5. RECAPITULATION AND DISCUSSION

The present work is part of a continued inquiry into the statistical nature and dynamics of seismogenetic systems. We address the question of whether seismogenesis is a Poisson (point) process in which successive failures are independent (*uncorrelated*) events, or a Complex process in which successive failures are dependent (*correlated*) due to fault-fault interaction over long spatiotemporal distances. Beyond its apparent academic interest, any answer to this problem might be useful in improving the forecasting of intermediate and long term earthquake hazard. Correlation confers memory through delayed action and manifests itself with power-law distributions of the temporal and spatial parameters of seismicity. We investigate the existence, intensity and evolution of correlation using the generalized formalism of Non Extensive Statistical Physics. In this context, the degree of non-equilibrium is summarized by the entropic index $q$ which is bounded in [0, 2], with $q=1$ corresponding to conservative Poissonian processes and $q>1$ indicating Complexity in non-conservative processes (Section 2.2). We focus on the analysis of two indices: The magnitude entropic index $q_M$ that is genetically related to the *b-value* of the Gutenberg-Richter law and the temporal entropic index $q_T$ that is associated with the statistics of the lapse between successive failures and indicates the extent of fault-fault interaction in a network. These are computed simultaneously by modelling bivariate empirical distributions of earthquake frequency vs. magnitude *and* interevent time.

Hitherto work by Efstathiou (2019), Efstathiou et al., (2015, 2016, 2017, 2018) and Tzanis et al., (2013, 2018) focused on the transformational plate boundaries of the north-eastern Circum-Pacific belt (California and Alaska, USA); only one convergent boundary was examined (Aleutian Arc). Herein, the emphasis is shifted to the convergent plate boundaries of the north-western Circum-Pacific belt: the Ryukyu Arc (RKU), the Izu–Bonin Arc (PSP) and the combined Okhotsk Plate, Honshu Arc and Pacific-Okhotsk plate boundary (OKH). The predominantly transformational continental tectonic domain of south-western Japan (SJP) is also considered. The analysis is based on the catalogue of the Japan Meteorological Agency for the period 2002–2016.5 (Section3); this contains both *crustal* earthquakes (in the cold brittle schizosphere) and *sub-crustal* earthquakes, mainly in Wadati-Benioff zones. In testing whether environmental (temperature, pressure) and boundary conditions (free at the surface, fixed otherwise) affect the statistical expression of seismogenesis, we separate crustal and sub-crustal earthquakes with respect to the Mohorovičić discontinuity and study them separately. Last but not least, we examine versions of the catalogue in which aftershocks were removed by stochastic declustering (Zhuang et al., 2002): if background seismicity is Poissonian, elimination of aftershocks should reduce the catalogue to a random set of events ($q_T \rightarrow 1$) otherwise the argument for Complexity would be stronger.

Our results comprise "expected" and "interesting" parts. "Expected" is the behaviour of the magnitude entropic indices $q_M$, which after conversion to $b_q$, (NESP generalization of the Gutenberg-Richter *b*-value) were found to be generally consistent with conventionally determined *b* values (Tables 2 and 3). This is important because the Gutenberg-Richter law *cannot* be derived from first principles in the context of point (Poisson) processes but *can* be derived in the generalized context of NESP. It follows that in terms of distribution and scaling, the fault networks of the north-western Circum-Pacific belt can be classified as *complex sub-* *extensive*. The magnitude entropic index is observed to be stable and consistent function of magnitude or/and interevent distance and to vary within a very limited range of values: herein this is 1.45 to 1.62. Naturally, it exhibits differences between seismogenetic systems but on average it is *comparable* between full crustal and sub-crustal processes, as well as between *background* crustal and sub-crustal processes. However, it is different between full and background processes. In the crust, the grand average of full processes is 1.56±0.032 and the grand average of background processes at the 80% probability level is 1.5±0.025. Below the crust, the corresponding grand averages are 1.54±0.02 and 1.5±0.026. The drop in $q_M$ corresponds to an increase in $b_q$ and can be explained as a consequence of relaxation in the clustering of faulting activity due to the removal of dependent events (aftershocks and swarms).

We now concentrate on the "interesting" part, i.e. the dynamics of seismogenesis. In doing that, let us first focus on the long-term correlation of *crustal* seismicity as characterized by the temporal entropic index $q_T$. With the exception of OKHC, full catalogues yield weak to moderate overall correlation. Definite increase of correlation with magnitude is observed in the transformational system of SJPC. Very notably, the same was observed in the transformational crustal systems of California and Alaska (Efstathiou et al., 2017, 2018; Tzanis et al., 2018). Steep increase of correlation with magnitude is also observed in OKH. In all cases, this is attributed to the progressive increase of the interaction radius (connectivity) associated with progressively larger earthquakes. Very strong correlation is generally observed at short (<100km) interevent distances, which can be easily explained as an effect of short-range interaction in aftershock and swarm clusters. Correlation levels drop to weak–moderate at interevent distances longer than 200km, although a low-rate but clear increase with interevent distance is observed. At ranges longer than 300km, correlation can hardly be explained by aftershocks: such interevent distances are significantly larger than the characteristic dimensions of aftershock zones associated with $M_w$ 7-7.3 earthquakes (e.g. Kagan, 2002). This implies a persistent degree of long-range interaction.

On declustering the catalogues, correlation drops to insignificant-weak in RKUC, PSPC and OKHC, but remains unchanged in the transformational SJPC. Increase of correlation with magnitude persists only in SJP, meaning that it was mainly due to interaction *within* earthquake clusters in OKHC. In terms of interevent distance, substantial short-range effects persist only in RKUC. Long-range correlation (interaction) remains weak to moderate and practically unaffected in all four systems, while low-rate increase with interevent distance also appears to persist. The persistence of correlation in the declustered catalogue of the transformational SJPC is consistent with analogous observations in the transformational plate boundaries of California and Alaska (Efstathiou, 2019; Efstathiou et al., 2017, 2018; Tzanis et al., 2018). Likewise, the significant decrease observed in the declustered crustal catalogues of convergent boundaries (RKUC, PSPC, OKHC) is consistent with analogous observations in the crustal seismicity of the Aleutian Arc (Efstathiou, 2019; Tzanis et al., 2018). It is thus tempting to suggest that the regional geodynamic setting is of significance to the development of Complexity.

With regard to long-term correlation in *sub-crustal* seismicity, full catalogues yield insignificant to very weak correlation both as function of magnitude and interevent distance. The temporal index will generally not exceed 1.15 except at short ranges where it is weak to moderate in RKUS, PSPS and OKHS, but very strong in SJPS. On declustering the catalogues one observes that the level of correlation either remains unchanged, or drops by a small amount. Moreover it reduces to insignificant-weak for short interevent distances at RKUS, PSPS and OKHS and cannot be evaluated in SJPS. It thus appears that in sub-crustal systems, the correlated component of seismicity is exclusively short-range and attributable to tenuous, short-lived aftershock sequences. Analogous results have been obtained for the Aleutian Wadati-Benioff zone (Efstathiou, 2019; Tzanis et al., 2018). Thus, the average (long-term) dynamics of sub-crustal systems appear to be *predominantly* (albeit not precisely) Poissonian.

It is interesting, to study the dynamic evolution of seismogenetic systems as imaged by the evolution of the temporal entropic index. In crustal full processes, very strong correlation is always observed immediately following the occurrence of large (M>6.8) events. This is a time/local effect that gradually diminishes as aftershock sequences die away and systems relaxes to states of lower correlation which is generally moderate. This is compatible with observations made in crustal seismogenetic systems of California and Alaska (Efstathiou, 2019; Efstathiou et al., 2015). Analogous observations can be made in regard to sub-crustal full processes albeit with two differences. First, the time-local increase of correlation after significant earthquakes is generally no higher than "strong" and decays faster than in the crust, apparently due to the tenuousness of sub-crustal aftershock sequences. Second, when in relaxed states correlation generally very weak to insignificant, i.e. *considerably* lower than that of crustal relaxed states!

In studying the evolution of correlation in background (global) processes, a general observation is that background correlated states are considerably lower than those of full processes. In the crust, $q_T$ appears to linger about 1.2 (weak); this contrasts corresponding results from California in which background correlation was found to *significantly increase* in comparison to the full processes (Efstathiou, 2019; Efstathiou et al., 2015). A very notable case is the Okhotsk plate system (OKH) in which correlation was practically nihil up to year 2009 and gradually increased to weak–moderate ahead of the 2011.19 Tōhoku event. Background correlation may vary dynamically: it is generally insignificant to weak but may rapidly transition to moderate for extended periods, ostensibly associated with temporal clusters of large earthquakes. Such transitions were observed *simultaneously* in the Wadati-Benioff zones of RKU, PSP and OKH and, *very conspicuously*, shortly before (onset) and after (offset) the 2011.19 Tōhoku mega-event (Fig. 12b and 15b respectively). For now, their origin cannot be specified but due to the spatial scales and "coincidences" it is reasonable to assume that it involves geodynamic factors, possibly transfer of stress during preparation of the Tōhoku earthquake. In this respect, OKH is of particular interest in that both crustal and sub-crustal background activity evolved from low to moderately correlated states ahead said event whose source area transited from quasi-Poissonian to moderately sub-extensive.

Based on the present and hitherto work one sees that different seismogenetic systems exhibit very different attributes of complexity. The crustal seismogenetic backgrounds of convergent plate boundaries appear to be weakly correlated and those of transformational plate boundaries moderately to very strongly correlated. At any rate, crustal systems appear to be Complex Sub-Extensive with some degree of long-range interaction. Conversely, sub-crustal systems and Wadati-Benioff zones verge on randomness albeit they may transition between states of higher and lower Complexity; long-range interaction is generally absent.

Consider, now, that Complexity may arise by several different mechanisms. Inasmuch as power-law distributions and long-range interaction are hallmarks of critical phenomena, the persistent correlation of crustal background processes could be construed as evidence of Criticality which, however, is neither uniform across the crust, *nor* stationary (i.e. of the SOC variant). On the other hand, Complexity and Criticality do not always go together and there are non-critical mechanisms that may generate power-laws (e.g. Sornette, 2004; Sornette and Werner, 2009), some of which may apply to sub-crustal seismicity, as will be discussed below.

We believe that we can formulate a basic interpretation of our observations based on fault networks with *small-world* topologies (e.g. Abe and Suzuki, 2004, 2007; Caruso et al., 2005, 2007). We are directed to this interpretation by the documented existence of long-range interaction, fruitful studies non-conservative small-world Olami-Feder-Christensen models (Caruso et al., 2005; Caruso et al., 2007), and suggestive evidence of small-worldliness in the seismicity of California and Japan (Abe and Suzuki, 2004, 2007). In such networks, each fault is a node that belongs to the hierarchy of some local cluster and interacts with proximal or distal nodes according to the connectivity and range of its hierarchical level. Upon excitation by some stress perturbation, a node responds by accumulating energy in the form of strain and transmitting/releasing it to *connected* nodes at various rates, thus operating as a delayed feedback loop and inducing heterogeneity in the distribution of stress transfer and release rates across the entire network; this appears to be important in the development of criticality in small-world networks (Yang, 2001; Caruso et al., 2007).

Consider, now, that crustal fault networks are subject to free boundary conditions at the Earth-Atmosphere interface: near-surface faults comprise boundary elements of the network. In OFC networks free boundary conditions force the boundary elements to interact at different (delayed) frequencies with respect to deeper buried elements and this induces partial synchronization of the boundary elements, building long-range spatial correlations and facilitating the development of a critical state (e.g. Lise and Paszucki; Caruso et al., 2005; Hergarten and Krenn, 2011). The particularly interesting study of Hergarten and Krenn (2011) indicates that the dynamics of a fault network may be gov-erned by two competing mechanisms: Synchronization, that pushes the system toward criticality and de-synchronization that prevents it from becoming overcritical and generates foreshocks and aftershocks. When the system has reached the critical state, synchronized failure transfers more stress to connected nodes and causes them to fail early, de-synchronizing with the rest of the system. When, however, the lag between de-synchronized failures becomes short again, the system can re-synchronize and repeat the cycle. This mechanism generates sequences of foreshocks, main shocks and aftershocks.

Based on the above considerations, it is plausible that the sub-extensivity of crustal fault networks is induced by the connectivity and synchronization of top-tier faults. In the transformational boundaries of California and Alaska, these are the contiguous segments of large transform fault zones that continuously "push" against each other and function as "hubs" that facilitate longitudinal interaction between remote clusters, maintaining the system in states of high correlation. In the crustal networks of convergent plate boundaries top-tier faults are essentially low-angle to sub-horizontal megathrusts, whose contiguous segments do *not* push against each and thus are not as strongly connected as large transform faults, maintaining the network in states of low correlation. This interpretation posits that free boundary conditions are central to the development of complexity and criticality. By inference, it also implies that the deep-seated fault networks of Wadati-Benioff zones would be kept away from Complexity as they are subject to fixed boundary conditions that inhibit synchronization.

Finally, it appears that deep-seated fault networks may transition between correlated and uncorrelated/ quasi-Poissonian states. With the limited evidence at hand, it is rather difficult to decide if the correlated states are generated internally (by self-organization), or induced externally and are therefore non-critical. As mentioned above there are complexity mechanisms that do not involve criticality, yet may maintain a fault system in a state of non-equilibrium; a list can be found in in Sornette (2004) and a comprehensive discussion in Sornette and Werner (2009). Celikoglu et al., (2010) applied the Coherent Noise Model (Newman, 1996) based on the notion of some external stress acting coherently onto all agents of the system without having any direct interaction with them. Although it has a weak point in that it does not consider the geometric configuration of the agents and how it would influence the behaviour of the system, the CNM was shown to generate power-law behaviour in interevent time distributions. At any rate, the fault networks of Wadati-Benioff zones appear to lack long-range interaction –hallmark of Criticality– and the transitions into and out of correlated states appear to be associated with the occurrence of major events. Accordingly, one might suggest that deep-seated networks, being "incapable" of synchronization and lacking crucial characteristics of criticality, may be driven to sub-extensivity by non-critical mechanisms such as external geodynamic forcing (as postulated in CNM). It follows that a lot of work is needed before sound conclusions can be reached.

## 6. ACKNOWLEDGEMENTS

We thank the National Research Institute for Earth Science and Disaster Resilience of Japan (NIED) and the Japan Meteorological Agency (JMA) for allowing us to use and publish data extracted from the JMA earthquake catalogue. The earthquake catalogue used in this study is produced by the Japan Meteorological Agency (JMA), in cooperation with the Ministry of Education, Culture, Sports, Science and Technology. The catalogue is based on data provided by the National Research Institute for Earth Science and Disaster Resilience, the Japan Meteorological Agency, Hokkaido University, Hirosaki University, Tōhoku University, the University of Tokyo, Nagoya University, Kyoto University, Kochi University, Kyushu University, Kagoshima University, the National Institute of Advanced Industrial Science and Technology, the Geographical Survey Institute, Tokyo Metropolis, Shizuoka Prefecture, Hot Springs Research Institute of Kanagawa Prefecture, Yokohama City, and Japan Agency for Marine-Earth Science and Technology.

**TABLE 1**

**Table 1.** Summary of the earthquake sub-catalogues used in the present analysis.

| Source Area | Source Area Code | Period | $M_c$ | Full catalogues | Declustered $\phi \geq 70\%$ | Declustered, $\phi \geq 80\%$ | Declustered, $\phi \geq 90\%$ |
|---|---|---|---|---|---|---|---|
| | | | | № events | № events | № events | № events |
| Crustal Seismicity of Ryukyu Arc. | **RKUC** | 2002-2016 | 3.0 | 6260 | 2352 | 2013 | 1654 |
| Sub-crustal Seismicity of Ryukyu Arc and subduction. | **RKUS** | 2002-2016 | 3.0 | 13430 | 8286 | 7556 | 6020 |
| Crustal Seismicity of Izu-Bonin Arc | **PSPC** | 2002-2016 | 3.0 | 2192 | 1221 | 1071 | 649 |
| Sub-crustal Seismicity of Izu-Bonin Arc and subduction. | **PSPS** | 2002-2016 | 3.0 | 13457 | 9065 | 7944 | 5004 |
| Crustal Seismicity of South-western Japan. | **SJPC** | 2002-2016 | 3.0 | 1763 | 851 | 823 | 766 |
| Sub-crustal Seismicity of South-western Japan. | **SJPS** | 2002-2016 | 3.0 | 1328 | 497 | 482 | 450 |
| Crustal Seismicity of Honshu Arc (Okhotsk Plate) | **OKHC** | 2002-2010.19 | 3.0 | 2918 | 957 | 912 | 792 |
| Sub-crustal Seismicity of Japan Trench and Pacific-Okhotsk subduction. | **OKHS** | 2002-2010.19 | 3.0 | 11514 | 6594 | 5732 | 3467 |

**TABLE 2**

**Table 2.** Summary of the entropic indices determined for *crustal* seismicity.

| | | $q_T(M_{th})$ Range | $\langle q_T(M_{th}) \rangle$ | $q_T(\Delta d)$ Range | | $q_M(M_{th})$ Range | $\langle q_M(M_{th}) \rangle$ | $\langle b_q(M_{th}) \rangle$ | $q_M(\Delta d)$ Range | $\langle q_M(\Delta d) \rangle$ | $\langle b_q(\Delta d) \rangle$ |
|---|---|---|---|---|---|---|---|---|---|---|---|
| | | | | $\Delta d \leq 100$km | $\Delta d > 100$km | | | | | | |
| **RKUC** | Full | 1.22-1.44 | 1.36±0.08 | 1.89–1.75 | 1.22-1.06 | 1.53-1.58 | 1.55±0.02 | 0.82±0.05 | 1.56-1.48 | 1.52±0.3 | 0.93±0.09 |
| | ≥ 70% | 1.13-1.22 | 1.17±0.03 | 1.77-1.23 | 1.38-1.12 | 1.49-1.46 | 1.48±0.01 | 1.11±0.04 | 1.49-1.44 | 1.46±0.02 | 1.16±0.08 |
| | ≥ 80% | 1.12-1.23 | 1.17±0.04 | 1.77 | 1.29-1.11 | 1.49-1.46 | 1.47±0.01 | 1.12±0.04 | 1.48-1.45 | 1.47±0.01 | 1.13±0.04 |
| | ≥ 90% | 1.14-1.25 | 1.19±0.04 | - | - | 1.50-1.47 | 1.48±0.01 | 1.07±0.05 | - | - | - |
| **PSPC** | Full | 1.14-1.28 | 1.23±0.05 | 1.96 | 1.37-1.09 | 1.56-1.58 | 1.56±0.01 | 0.78±0.02 | 1.62-1.55 | 1.59±0.02 | 0.69±0.07 |
| | ≥ 70% | 1.09-1.22 | 1.15±0.04 | 1.16 | 1.14-1.36 | 1.57-1.51 | 1.53±0.02 | 0.88±0.06 | 1.62-1.57 | 1.59±0.02 | 0.7±0.06 |
| | ≥ 80% | 1.07-1.22 | 1.14±0.04 | 1.21 | 1.30-1.16- | 1.57-1.51 | 1.53±0.02 | 0.88±0.06 | 1.55-1.60 | 1.59±0.02 | 0.7±0.06 |
| **SJPC** | Full | 1.18-1.51 | 1.30±0.13 | 1.91 | 1.18-1.12 | 1.52-1.55 | 1.53±0.01 | 0.90±0.03 | 1.54-1.52 | 1.53±0.01 | 0.9±0.03 |
| | ≥ 70% | 1.09-1.34 | 1.24±0.1 | 1.31 | 1.16-1.07 | 1.5-1.48 | 1.49±0.01 | 1.05±0.03 | 1.50-1.49 | 1.49±0.01 | 1.03±0.02 |
| | ≥ 80% | 1.09-1.33 | 1.23±0.11 | - | - | 1.5-1.47 | 1.49±0.01 | 1.05±0.04 | - | - | - |
| **OKHC** | Full | 1.2-1.89 | 1.49±0.24 | 1.81 | 1.32-1.18 | 1.55-1.60 | 1.58±0.02 | 0.72±0.05 | 1.59-1.53 | 1.56±0.02 | 0.79±0.05 |
| | ≥ 70% | 1.05-1.13 | 1.1±0.03 | - | 1.37-1.25 | 1.51-1.497 | 1.50±0.004 | 0.99±0.02 | 1.54-1.51 | 1.52±0.01 | 0.93±0.05 |
| | ≥ 80% | 1.05-1.1 | 1.07±0.02 | - | 1.34-1.18 | 1.51-1.50 | 1.50±0.004 | 0.98±0.02 | 1.52-1.52 | 1.52±0.01 | 0.92±0.03 |
| | ≥ 90% | 1.05-1.12 | 1.09±0.04 | - | - | 1.50 | 1.50±0.001 | 1.0±0.004 | - | - | - |

**TABLE 3**

**Table 3.** Summary of entropic indices determined for *sub-crustal* seismicity.

| | | $q_T(M_{th})$ Range | $\langle q_T(M_{th})\rangle$ | $q_T(\Delta d)$ Range | | $q_M(M_{th})$ Range | $\langle q_M(M_{th})\rangle$ | $\langle b_q(M_{th})\rangle$ | $q_M(\Delta d)$ Range | $\langle q_M(\Delta d)\rangle$ | $\langle b_q(\Delta d)\rangle$ |
|---|---|---|---|---|---|---|---|---|---|---|---|
| | | | | $\Delta d \leq 100$km | $\Delta d > 100$km | | | | | | |
| RKUS | Full | 1.08-1.21 | 1.13±0.035 | 1.32-1.12 | 1.13-1.03 | 1.55-1.52 | 1.53±0.01 | 0.87±0.03 | 1.59-1.51 | 1.54±0.02 | 0.84±0.07 |
| | ≥ 70% | 1.01-1.15 | 1.08±0.04 | 1.11-1.12 | 1.14-1.04 | 1.52-1.50 | 1.51±0.01 | 0.97±0.03 | 1.56-1.50 | 1.52±0.02 | 0.91±0.06 |
| | ≥ 80% | 1.03-1.14 | 1.09±0.033 | 1.17 | 1.15-1.06 | 1.52-1.49 | 1.51±0.01 | 0.97±0.04 | 1.54-1.51 | 1.52±0.01 | 0.91±0.04 |
| | ≥ 90% | 1.00-1.14 | 1.08±0.038 | 1.06 | 1.15-1.06 | 1.52-1.49 | 1.51±0.01 | 0.98±0.04 | 1.54-1.50 | 1.52±0.01 | 0.93±0.05 |
| PSPS | Full | 1.00-1.20 | 1.08±0.049 | 1.44-1.20 | 1.12-1.06 | 1.55-1.51 | 1.53±0.01 | 0.90±0.05 | 1.57-1.53 | 1.55±0.01 | 0.83±0.05 |
| | ≥ 70% | 1.00-1.21 | 1.09±0.067 | 1.25-1.108 | 1.15-1.08 | 1.52-1.47 | 1.50±0.02 | 1.0±0.07 | 1.59-1.53 | 1.56±0.02 | 0.78±0.06 |
| | ≥ 80% | 1.00-1.18 | 1.08±0.054 | 1.22-1.20 | 1.2-1.04 | 1.54-1.45 | 1.50±0.02 | 1.0±0.1 | 1.58-1.54 | 1.56±0.01 | 0.79±0.04 |
| | ≥ 90% | 1.03-1.20 | 1.09±0.057 | 1.27 | 1.14-1.03 | 1.56-1.49 | 1.52±0.02 | 0.93±0.07 | 1.6-1.55 | 1.57±0.02 | 0.75±0.05 |
| SJPS | Full | 1.03-1.18 | 1.11±0.06 | 1.74 | 1.40-1.12 | 1.55-1.51 | 1.54±0.01 | 0.87±0.04 | 1.59-1.53 | 1.55±0.03 | 0.82±0.09 |
| | ≥ 70% | 1.00-1.15 | 1.06±0.065 | | | 1.54-1.52 | 1.52±0.01 | 0.92±0.03 | | | |
| OKHS | Full | 1.07-1.25 | 1.16±0.038 | 1.30-1.18 | 1.17-1.09 | 1.57-1.55 | 1.56±0.01 | 0.79±0.03 | 1.59-1.54 | 1.57±0.02 | 0.77±0.06 |
| | ≥ 70% | 1.06-1.2 | 1.12±0.039 | 1.18-1.14 | 1.175-1.05 | 1.51-1.48 | 1.50±0.01 | 1.01±0.03 | 1.56-1.51 | 1.54±0.02 | 0.87±0.06 |
| | ≥ 80% | 1.08-1.19 | 1.14±0.03 | 1.26-1.11- | 1.22-1.03 | 1.5-1.47 | 1.49±0.01 | 1.05±0.03 | 1.57-1.52 | 1.54±0.02 | 0.87±0.06 |
| | ≥ 90% | 1.01-1.12 | 1.09±0.03 | 1.03 | 1.22-1.02 | 1.50-1.48 | 1.49±0.01 | 1.03±0.02 | 1.56-1.52 | 1.54±0.01 | 0.84±0.03 |